\documentclass[twocolumn,english,aps,prd,reprint,floatfix,notitlepage,footinbib,preprintnumbers,superscriptaddress,altaffilletter]{revtex4-2}

\usepackage{amsmath,amssymb,amsfonts}
\usepackage{hyperref,breakurl,cleveref,url}
\usepackage{color}

\usepackage{graphicx}
\usepackage[export]{adjustbox}

\usepackage{accents,mathrsfs,mathtools}

\renewcommand{\paragraph}[1]{%
	\textit{#1}.---%
}

\def\skip{\vskip1.5pt}
\newcommand\trick[1]{}

\usepackage{enumitem}
\setlist[enumerate]{
	label={},
	leftmargin=2em,
	itemsep=2pt,
	topsep= 2pt,
	partopsep=0pt,
	parsep=0pt,
}

\let\oldeqref\eqref
\renewcommand{\eqref}[1]{Eq.\,\smash{\oldeqref{#1}}}
\newcommand{\eqrefs}[2]{Eqs.\,\smash{\oldeqref{#1}} and \smash{\oldeqref{#2}}}

\newcommand{\rcite}[1]{Ref.\,\cite{#1}}
\newcommand{\rrcite}[1]{Refs.\,\cite{#1}}

\newcommand{\fref}[1]{Fig.\,\ref{#1}}

\newcommand{\App}[1]{Appendix~\ref{#1}}

\def\mem{\hspace{0.1em}}
\def\hem{\hspace{0.05em}}
\def\nem{\hspace{-0.1em}}
\def\hnem{\hspace{-0.05em}}
\def\hhem{\hspace{0.025em}}

\def\blank{{\,\,\,\,\,}}

\def\qiq{{\quad\implies\quad}}

\def\a{\alpha}

\def\c{{\gamma}}

\def\e{\epsilon}

\def\m{\mu}
\def\n{\nu}
\def\r{\rho}
\def\s{\sigma}
\def\k{\kappa}
\def\l{\lambda}
\def\t{\tau}

\def\bpsi{\bar{\psi}}

\def\bQ{\bar{Q}}

\def\mtimes{{\mem\times\mem}}

\newcommand{\BB}[1]{\Big(\,{#1}\,\Big)}
\newcommand{\bb}[1]{\bigg(\,{#1}\,\bigg)}
\newcommand{\bigbig}[1]{\big(\mem{#1}\mem\big)}

\newcommand{\wrap}[1]{{\smash{#1}\vphantom{\beta}}}

\def\lsq{{
		\kern-0.037em
		\adjustbox{scale=0.919,valign=c}{$
			{
				\adjustbox{raise=-0.0855em}{$\lfloor$}
				\llap{\reflectbox{\rotatebox[origin=c]{180}{$\lfloor$}}}
			}
			$}
		\kern-0.04em
}}
\def\rsq{{
		\kern-0.04em
		\adjustbox{scale=0.919,valign=c}{$
			{
				\rlap{\reflectbox{\rotatebox[origin=c]{180}{$\rfloor$}}} 
				\adjustbox{raise=-0.0855em}{$\rfloor$}
			}
			$}
		\kern-0.037em
}}

\usepackage{hyphenat}
\usepackage{lipsum}

\usepackage{simpler-wick}

\newcommand{\Jac}[1]{\mathrm{Jac}\hem\big(\hem{#1}\hem\big)}

\def\lb{\{\kern-0.15em\{}
\def\rb{\}\kern-0.15em\}}

\newcommand{\pb}[2]{\{\hem{#1},{#2}\hem\}}
\newcommand{\dpb}[2]{\lb\hem{#1},{#2}\hem\rb}

\newcommand{\comm}[2]{[\hem{#1},{#2}\hem]}

\def\R{\mathbb{R}}
\def\g{\mathfrak{g}}
\def\su{\mathfrak{su}}
\def\diff{\mathfrak{diff}}
\def\SU{\mathrm{SU}}

\def\P{\mathcal{P}}
\def\Aflat{\mathbb{A}}
\def\A{\mathcal{A}}

\def\can{{\text{can}}}
\def\cov{{\adjustbox{raise=-0.06em,scale=0.85}{${}_\nabla$}}}

\def\YM{{\text{YM}}}

\def\O{\mathcal{O}}
\def\N{\mathcal{N}}

\def\M{\mathcal{M}}

 \def\mwedge{{\mem\wedge\mem\hhem}}
\def\swedge{{\mem{\wedge}\,}}
\def\mtimes{{\mem\times\mem}}

\def\mplus{{\mem+\mem}}

\def\tensor\otimes

\def\bplus{{\,+\,}}

\usepackage{tikz}
\usetikzlibrary{calc} 
\usetikzlibrary{shapes.geometric} 
\usetikzlibrary{positioning} 
\usetikzlibrary{fit} 
\usepackage[a]{esvect} 
\tikzset{empty/.style = {inner sep = 0pt, outer sep = 0, minimum size = 0}}
\tikzset{b/.style = {inner sep = 2pt, outer sep = 4pt, minimum size = 12pt}}
\tikzset{w/.style = {inner sep = 1pt, outer sep = 2pt, minimum size = 12pt, anchor = west}}
\usepackage[export]{adjustbox}
\tikzset{every node/.style = {inner sep = 0pt, outer sep = 0, minimum size = 0}}

\tikzset{dot/.style = {circle, draw=black, fill=black, inner sep=0pt, outer sep=0pt, minimum size=2.5pt, line width=1.2pt}}

\definecolor{lgray}{RGB}{150,150,150}
\tikzset{
	dprop/.style = {
		draw, line width=0.8pt,
		dotted, 
		line cap=round,
		dash pattern=on 0pt off 2.53pt,
		color=lgray
	}
}

\tikzset{l/.style = {draw, line width = 1.2pt}}

\tikzset{s/.style = {inner sep = 2.5pt, outer sep =2.5pt, minimum size = 1pt, font = \small}}

\newcommand{\hlc}[1]{{\color[RGB]{0,145,225}{}#1}}

\newcommand{\hld}[1]{{\color[RGB]{130,100,40}{}#1}}

\newcommand{\hlg}[1]{{\color[RGB]{2,145,15}{}#1}}

\newcommand{\hlx}[1]{{\color[RGB]{220,10,10}{}#1}}

\definecolor{guides}{RGB}{143,153,173}

\newcommand{\contd}[4]{{\color{guides}\small%
	\text{\quad\, from \quad} 
		\bigg\{
		\begin{aligned}[c]
			#1 &\text{---} #2
			\\[-0.4\baselineskip]
			#3 &\text{---} #4
		\end{aligned} 
		\bigg\}
}}

\newcommand{\contx}[4]{
\times 2
{\color{guides}\small%
	\text{\quad\, from \quad}
		\bigg\{
		\begin{aligned}[c]
			#1 &\text{---} #2
			\\[-0.4\baselineskip]
			#3 &\text{---} #4
		\end{aligned} 
		\bigg\}
	\text{\,,\,\,}
		\bigg\{
		\begin{aligned}[c]
			#3 &\text{---} #4
			\\[-0.4\baselineskip]
			#1 &\text{---} #2
		\end{aligned} 
		\bigg\}
	\,,
}}
\newcommand{\contxd}[4]{
\times 2
{\color{guides}\small%
	\text{\quad\, from \quad}
		\bigg\{
		\begin{aligned}[c]
			#1 &\text{---} #2
			\\[-0.4\baselineskip]
			#3 &\text{---} #4
		\end{aligned} 
		\bigg\}
	\text{\,,\,\,}
		\bigg\{
		\begin{aligned}[c]
			#3 &\text{---} #4
			\\[-0.4\baselineskip]
			#1 &\text{---} #2
		\end{aligned} 
		\bigg\}
}}

\newcommand{\conty}[4]{
\times 2
{\color{guides}\small%
	\text{\quad\, from \quad}
		\bigg\{
		\begin{aligned}[c]
			#1 &\text{---} #2
			\\[-0.4\baselineskip]
			#3 &\text{---} #4
		\end{aligned} 
		\bigg\}
	\text{\,,\,\,}
		\bigg\{
		\begin{aligned}[c]
			#2 &\text{---} #1
			\\[-0.4\baselineskip]
			#4 &\text{---} #3
		\end{aligned} 
		\bigg\}
	\,,
}}
\newcommand{\contyd}[4]{
\times 2
{\color{guides}\small%
	\text{\quad\, from \quad}
		\bigg\{
		\begin{aligned}[c]
			#1 &\text{---} #2
			\\[-0.4\baselineskip]
			#3 &\text{---} #4
		\end{aligned} 
		\bigg\}
	\text{\,,\,\,}
		\bigg\{
		\begin{aligned}[c]
			#2 &\text{---} #1
			\\[-0.4\baselineskip]
			#4 &\text{---} #3
		\end{aligned} 
		\bigg\}
}}

\newcommand{\contxy}[4]{
\times 4
{\color{guides}\small%
	\text{\quad\, from \quad}
		\bigg\{
		\begin{aligned}[c]
			#1 &\text{---} #2
			\\[-0.4\baselineskip]
			#3 &\text{---} #4
		\end{aligned} 
		\bigg\}
	\text{\,,\,\,}
		\bigg\{
		\begin{aligned}[c]
			#2 &\text{---} #1
			\\[-0.4\baselineskip]
			#4 &\text{---} #3
		\end{aligned} 
		\bigg\}
	\text{\,,\,\,}
		\bigg\{
		\begin{aligned}[c]
			#3 &\text{---} #4
			\\[-0.4\baselineskip]
			#1 &\text{---} #2
		\end{aligned} 
		\bigg\}
	\text{\,,\,\,}
		\bigg\{
		\begin{aligned}[c]
			#4 &\text{---} #3
			\\[-0.4\baselineskip]
			#2 &\text{---} #1
		\end{aligned} 
		\bigg\}
	\,,
}}

\def\X{\mathbf{X}}
\def\P{\mathbf{P}}

\def\Th{\mathbf{\Theta}}
\def\bTh{\bar{\mathbf{\Theta}}}

\def\Ps{\mathbf{\Psi}}
\def\bPs{\bar{\mathbf{\Psi}}}

\def\E{\mathbf{E}}

\def\Ric{{\mathrm{Ric}}}

\newcommand{\LA}[1]{{\smash{
	\accentset{\footnotesize\blacktriangleleft}{#1}
}}}
\newcommand{\RA}[1]{{\smash{
	\accentset{\footnotesize\blacktriangleright}{#1}
}}}

\def\LPs{\LA{\mathbf{\Psi}}}

\def\RPs{\RA{\mathbf{\Psi}}}
\def\RbPs{\RA{\bar{\mathbf{\Psi}}}}

\def\LTh{\LA{\mathbf{\Theta}}}

\def\RbTh{\RA{\bar{\mathbf{\Theta}}}}

\newcommand{\la}[1]{{\smash{
	\accentset{\adjustbox{scale=0.45}{$\scriptsize\blacktriangleleft$}}{#1}
}}}
\newcommand{\ra}[1]{{\smash{
	\accentset{\adjustbox{scale=0.45}{$\scriptsize\blacktriangleright$}}{#1}
}}}

\def\lPs{\la{\mathbf{\Psi}}}

\def\rbPs{\ra{\bar{\mathbf{\Psi}}}}

\usepackage{makecell}

\newcommand{\para}[1]{\noindent\textbf{#1.}|}

\def\mflat{\mathbb{M}}
\def\V{\mathbb{V}}

\def\btheta{\bar{\theta}}

\def\YM{{\text{YM}}}

\def\Grav{{\text{Grav}}}

\newcommand{\normal}[1]{{{:}\mem{#1}\mem{:}}}

\def\ta{{\smash{\tilde{a}}}{}}
\def\tb{{\smash{\tilde{b}}}{}}
\def\tc{{\smash{\tilde{c}}}{}}
\def\td{{\smash{\tilde{d}}}{}}

\def\tf{{\smash{\tilde{f}}}{}}

\def\R{\mathbb{R}}
\def\C{\mathbb{C}}

\def\so{\mathfrak{so}}

\def\g{\mathfrak{g}}  
\def\tg{\tilde{\mathfrak{g}}}

\def\diff{\mathfrak{diff}}  
\def\sdiff{\mathfrak{sdiff}}

\begin{document}
	
	\title{
		Double Copy and the Double Poisson Bracket
	}
	
	\author{Joon-Hwi Kim}
	\affiliation{Walter Burke Institute for Theoretical Physics, California Institute of Technology, Pasadena, CA 91125, USA}
	
	\begin{abstract}
		We derive 
		first-order and second-order
		field equations 
		from ambitwistor spaces as phase spaces of massless particles.
		In particular,
		the second-order field equations
		of Yang-Mills theory and general relativity
		are formulated
		in a unified form $\dpb{H}{H}_\cov = 0$,
		whose left-hand side
		describes a doubling of Poisson bracket
		in a covariant sense.
		This structure
		originates from a one-loop diagram
		encoded in gauge-covariant, associative operator products
		on the ambitwistor worldlines.
		A conjecture arises that
		the kinematic algebra might manifest 
		as the Poisson algebra
		of ambitwistor space.
	\end{abstract}
	
	\preprint{CALT-TH 2025-005}
	
	
	\bibliographystyle{utphys-modified}
	
	\renewcommand*{\bibfont}{\fontsize{8}{8.5}\selectfont}
	\setlength{\bibsep}{1pt}
	
	\maketitle
	
	\paragraph{Introduction}%
In string theory,
one unlocks 
unique perspectives toward
the dynamics of fields and spacetime.
A textbook result
is the derivation of vacuum Einstein's equations
from vanishing Weyl anomaly on the worldsheet
\cite{Alvarez-Gaume:1980zra,Alvarez-Gaume:1981exa,Callan:1985ia}.
Intuitively, one may picture the string as a picky entity, 
demanding specific conditions on the background in which it seeks to reside.
This provides a prototype of the idea that
field equations can arise as
consistency conditions
imposed by test objects
\cite{Alvarez-Gaume:1980zra,Alvarez-Gaume:1981exa,Callan:1985ia,Abouelsaood:1986gd,Banks:1986fu}.

A particle, on the other hand,
is seemingly an object far less pickier than the string.
However, according to 
Feynman's view
\cite{dyson1990feynman},
a particle still demands a condition on the background it couples to,
via
gauge covariance and associativity of its quantum-mechanical operator algebra.

Suppose a charged particle is put in
a background electromagnetic field $F_{mn}$.
A manifestly gauge-invariant formulation of its quantum mechanics
is viable by employing noncanonical commutation relations
\cite{souriau1970structure,dyson1990feynman}:
$[\hat{x}^m,\hat{x}^n] = 0$,
$[\hat{x}^m,\hat{p}_n] = i\hbar\mem \delta^m{}_n$,
and
$[\hat{p}_m,\hat{p}_n] = F_{mn}(\hat{x})$.
In this setup,
Feynman \cite{dyson1990feynman} imposes 
the Jacobi identity,
\begin{align}
	\label{!magneticQ}
	\comm{
		\comm{\hem \hat{p}_\wrap{[m}}{\hat{p}_\wrap{n}}
	}{
		\hat{p}_\wrap{r]}
	}
	\,=\ 
		0
	\,,
\end{align}
as an implication of 
having an associative operator product.
Notably, this derives a half of Maxwell's equations,
$\partial_\wrap{[r} F_\wrap{mn]} = 0$.

The above argument due to Feynman
has been applied to various kinds of particles
throughout works \cite{lee1990feynman,stern1993deformed,Tanimura:1992gn,chou1994dynamical,soni1992classical}.
As a result,
the field equations of the magnetic (Bianchi) type
are derived for nonabelian gauge theory and gravity
by imposing \eqref{!magneticQ}.
However, the electric-type field equations are missing.
For instance,
how could one derive 
the other half
of the Maxwell's equations,
$\partial^n F_{mn} = 0$?

Regarding this inquiry,
we may recall
a 1989 paper
by Mason and Newman 
\cite{mason-newman},
the title of which reads suggestive in a modern context:
``A connection between the Einstein and Yang-Mills equations.''
There, the authors explore the idea of 
identifying the electric-type field equations 
in terms of commutators of covariant derivatives.
In our particle perspective,
this translates to the following equation
because the kinetic momentum $\hat{p}_m$
describes the generator of gauge-covariant translations:
\begin{align}
	\label{!electricQ}
	\comm{
		\comm{\hem\hat p_m}{\hat p_n}
	}{
		\hat p^n
	}
	\,=\ 
		0
	\,.
\end{align}
Certainly, \eqref{!electricQ} derives 
the electric equations
$\partial^n F_{mn} = 0$
for Maxwell theory. 
In the same way,
Mason and Newman \cite{mason-newman} shows that
(in their language)
the Yang-Mills (YM) equations
arise also from \eqref{!electricQ}.
Unfortunately, however,
their attempt toward 
general relativity
does not end in full satisfaction.
Moreover, 
no clear physical origin was
identified
for the postulate in \eqref{!electricQ}.

At this moment,
we fast-forward the historical timeline to 
2010s
and witness that the field equations of 
YM theory and gravity
are derived 
in a worldsheet model \cite{adamo-ym,adamo-gravity}:
ambitwistor strings.
Ambitwistor strings are closed \textit{chiral} strings endowed with various matter content
\cite{Mason:2013sva,Geyer:2022cey}.
This construction 
has explained
the Cachazo-He-Yuan formulae \cite{Cachazo:2013gna,Cachazo:2013hca},
which
represent scattering amplitudes of massless particles
as moduli space integrals of 
a factorized integrand.
Notably,
this factorization derives concrete expressions to 
the Kawai-Lewellen-Tye 
\cite{KLT} version of double copy
\cite{Geyer:2022cey}.
Crucially,
\rrcite{adamo-ym,adamo-gravity} have established
that the magnetic and electric first-order field equations of YM theory
and the NS-NS sector of type II supergravity
can be derived by demanding a quantum consistency condition
on the ambitwistor worldsheet
\cite{Geyer:2022cey}.

In this paper, three key observations are made.
First,
the computations in \rrcite{adamo-ym,adamo-gravity}
could in fact be effectively performed 
in 
ambitwistor spaces
as \textit{particle} phase spaces,
by virtue of the very chiral nature of ambitwistor strings.
Second, 
attention should be given to
the \textit{second-order} field equations
that follow within the same framework,
which are not explicitly worked out
at least
in \rrcite{adamo-ym,adamo-gravity}.
Third,
the one-loop part of
the quantum-mechanical operator algebra
yields a \textit{doubling of Poisson bracket}
when described in terms of 
the Moyal star product.
The first insight may have been available 
from
\rrcite{Shi:2021qsb,Bonezzi:2025bgv,Bonezzi:2024emt,Bonezzi:2020jjq},
but the latter two seem not clearly realized.

Consequently,
we formulate
YM theory and gravity
in a unified fashion
from ambitwistor worldlines.
Notably, the resulting first-order equations 
identify a physical origin for
the structures 
foreshadowed in the classic era,
i.e., \eqrefs{!magneticQ}{!electricQ},
in terms of a supersymmetry.
Moreover, the second-order equations are formulated as
\begin{align}
	\label{!2}
	\dpb{H}{H}_\cov \,=\, 0
	\,.
\end{align}
Here, $\dpb{\blank}{\blank}_\cov$ is a bi-differential operator
that describes the doubling of Poisson bracket in a covariant sense,
while $H = \frac{1}{2}\mem p^2 + \cdots$
is the deformed mass-shell constraint
of the curved ambitwistor space.
\eqref{!2} computes a one-loop diagram encoded in a gauge-covariant associative operator algebra,
the constructive existence of which we prove by the Fedosov \cite{fedosov1994simple} theory.

This leads to
a conjecture
that 
the Bern-Carrasco-Johansson (BCJ) \cite{BCJ1, BCJ2} form of double copy
might also be derived from ambitwistor space:
\begin{align}
	\label{conjecture}
	\Box\mem V \,=\, \tfrac{1}{2}\mem \dpb{V}{V}
	\,.
\end{align}
\eqref{conjecture} 
briefly sketches this idea,
which arises by
splitting $H$ in \eqref{!2} into
$\frac{1}{2}\mem p^2$ and a ``vertex operator'' $V$.
The former
converts to the Laplacian $\Box$,
which is a \textit{second-order} operator.
%
%
%
Crucially,
\eqref{conjecture} is a case of the bi-adjoint scalar (BAS) equation,
which has served as the universal grammar for 
established instances of manifest BCJ duality:
the heavenly equation 
for self-dual gravity
\cite{DonalSDYM1,DonalSDYM2,DonalSDYM3,HenrikSDYM,plebanski1975some,park1990self,husain1994self,plebanski1994self,Plebanski:1995gk}
and special galileon in two dimensions
\cite{flatland,Zakharov:1973pp,Armstrong-Williams:2022apo}
all describe \eqref{conjecture}
with well-studied double Poisson brackets.
If \eqref{!2} can be converted to \eqref{conjecture},
the kinematic algebra of gravity
(general dimensions, non-self-dual)
will
manifest as
the Poisson algebra of ambitwistor space.

\skip
\paragraph{Classical Mechanics in Phase Space}%
Let us begin by describing
the old ideas of Feynman-Souriau \cite{dyson1990feynman,souriau1970structure}
and Mason-Newman \cite{mason-newman}.
For the sake of precision,
we first specialize in \textit{classical} Hamiltonian mechanics,
in which case 
the conditions in \eqrefs{!magneticQ}{!electricQ}
boil down to
the following 
classical 
avatars in phase space:
\begin{align}
	\label{!magnetic}
	\pb{
		\pb{p_\wrap{[m}}{p_\wrap{n}}
	}{
		p_\wrap{r]}
	}
	\,=\ 
		0
	\,,\\
	\label{!electric}
	\pb{
		\pb{p_m}{p_n}
	}{
		p^n
	}
	\,=\ 
		0
	\,.
\end{align}
Note that the Jacobi identity of the Poisson bracket
encodes
a principle in classical mechanics that time evolution preserves the Poisson bracket.

\skip
\paragraph{YM Theory from Colored Scalar Particle}%
Concretely, let us show how YM theory can be derived from 
\eqrefs{!magnetic}{!electric}.
This revisits the analysis in 
\rrcite{lee1990feynman,stern1993deformed,Tanimura:1992gn}.

Suppose a scalar particle in $d$-dimensional flat spacetime,
carrying a color charge $q_a$ valued in the dual of
the Lie algebra $\g = \su(N)$.
The particle's phase space can be realized as a Poisson manifold 
coordinatized by
$(x^m,p_m) \in T^*\R^d$
and
$q_a \in \g^*$.
In the free theory,
this phase space features
the nonvanishing Poisson brackets
\begin{align}
	\label{xpq-free}
	\pb{x^m}{p_n}
	\,=\,
		\delta^m{}_n
	\,,\quad
	\pb{q_a}{q_b}
	\,=\,
		q_c\mem f^c{}_{ab}
	\,,
\end{align}
where $f^c{}_{ab}$ are the structure constants.

To couple this particle to external fields
in a manifestly gauge-covariant fashion,
one modifies the Poisson structure in phase space.
This insight is due to
Feynman \cite{dyson1990feynman}
(in the Poisson language)
and
Souriau \cite{souriau1970structure}
(in the symplectic language).
A generic modification that preserves
the $x$-$x$ and $q$-$q$ brackets
is given in the form
\begin{align}
\begin{split}
	\label{YM.feynbr}
	\pb{q_a}{p_m}
	&\,=\,
		q_b\mem f^b{}_{ca}\mem A^c{}_m(x)
	\,,\\
	\pb{p_m}{p_n}
	&\,=\,
		q_a\mem F^a{}_{mn}(x)
	\,,
\end{split}
\end{align}
where $A$ and $F$ are introduced as independent fields.

To proceed, we evaluate \eqref{!magnetic}
with the brackets in \eqrefs{xpq-free}{YM.feynbr}.
This yields
\begin{align}
	\label{YM.1m}
	D_\wrap{[r} F^a{}_\wrap{mn]}
	\,=\,
		0
	\,,
\end{align}
namely the magnetic-type field equations of nonabelian gauge theory.
Here, $D$ denotes the covariant derivative
using $A$ as the gauge connection.

Thus,
we find that
a nonabelian gauge field
is coupled to the particle,
though its dynamics is not fully specified.
To this end, we impose
the postulate in \eqref{!electric}
and obtain
\begin{align}
	\label{YM.1e}
	D^n F^a{}_\wrap{mn}
	\,=\,
		0
	\,,
\end{align}
which are precisely the electric-type equations in YM theory.
This specifies that the nonabelian gauge theory
coupled to the particle
is YM theory, in particular.

In sum,
we have derived YM theory
from a classical colored scalar particle
via \eqrefs{!magnetic}{!electric}.

Geometrically,
our phase space approach has recast the gauge-covariant derivative $D_m$ in \rcite{mason-newman}
as the Hamiltonian vector field $\pb{\blank}{p_m}$ of the kinetic momentum $p_m$.

\paragraph{Gauge Covariance Versus Associativity: Classical}%
The fact that $x^m,p_m,q_a$
are all gauge-covariant variables
is easily checked by 
reproducing the Wong's equations \cite{wong1970field} as
the Hamiltonian equations of motion,
for instance.
Especially,
the \textit{kinetic} momentum $p_m$ equals the physical velocity $\dot{x}^m$ upon index raising.
Thus, the description of the particle with $x^m,p_m,q_a$
manifests gauge covariance.

In contrast, 
one can also employ
the \textit{canonical} \footnote{
	A caveat here is that the term ``canonical'' usually presumes
	\textit{symplectic} geometry,
	meaning Darboux coordinates \cite{darboux1882probleme}.
} momentum $p_m^\can = p_m + q_a A^a{}_m(x)$.
In this case, the particle's Poisson brackets are 
kept the same,
so the Jacobi identity is trivialized
from the free theory.
Thus, the description of the particle with $x^m,p_m^\can,q_a$
manifests 
that classical time evolution preserves the Poisson bracket.

This analysis 
reveals
a tension between 
two principles in classical mechanics.
The kinetic momentum is gauge covariant 
but makes Jacobi identity nontrivial,
facilitating the Feynman derivation.
The canonical momentum is not gauge covariant 
but trivializes Jacobi identity.
In fact,
this demonstrates
a classical vestige of
the tension between 
gauge covariance and associativity
in the quantum operator algebra,
which will be explored later.

\skip
\paragraph{Historical Remarks}%
In Sections 5 and 6 of \rcite{mason-newman},
Mason and Newman
attempt to derive general relativity
also from their postulate.
However,
a simple construction based on a scalar particle
yields instead
an alternative theory based on teleparallelism \cite{mason-newman,cartan1979letters}.
In fact,
we remark that
this teleparallel theory
is
Born-Infeld theory \cite{born1935quantization}
in disguise,
due to a modern reformulation \cite{cck}.

Regarding this failure,
Mason and Newman speculate that
a missing element might be spin (local Lorentz generators).
According to our modern understanding,
this indeed was a reasonable guess:
the particle one couples to the backgrounds
could be interpreted as the massless excitation itself,
and gluons and gravitons do carry spin.

With these remarks made,
we now switch to the modern constructions
as promised,
which are
based on the classical and quantum geometry of ambitwistor space.

\skip
\paragraph{Ambitwistor Space as a Constrained Phase Space}%
We begin with a friendly introduction to ambitwistor space.
The definition of
ambitwistor space is the space of complexified null geodesics
\footnote{Scaled, to be precise.}.
Physically,
it can be realized as
a constrained phase space
(symplectic quotient)
for an on-shell massless particle.
It is often extended by extra variables encoding color or spin.
See \rcite{Mason:2013sva} for a systematic exposition.

We shall present a concrete example.
Let $\mflat {\,=\,} (\R^d,\eta)$ be flat spacetime.
Consider the space
\begin{align}
	\label{Aflat-YM}
	\Aflat_\YM
	\,=\,
		T^*\mflat \mtimes \Pi\mflat \mtimes T^*\Pi\V
	\,,
\end{align}
coordinatized by bosonic variables
$(x^m,p_m) \in T^*\R^d$
and fermionic variables
$\psi^m \in \Pi\R^d$,
$\theta^i \in \Pi\V$,
$\btheta_i \in \Pi\V^*$.
Here, we have supposed a representation $\rho: G \to \mathrm{GL}(\V)$ of
the Lie group $G = \SU(N)$,
where
$\V$ is a vector space
assigned with indices $i,j,k,l,\cdots$.

$\Aflat_\YM$ is a symplectic manifold that
serves as the phase space for an off-shell colored spinning particle.
In the free theory, 
the symplectic form is $\omega = dp_m \swedge dx^m + \frac{i}{2}\mem \eta_{mn}\mem d\psi^m \swedge d\psi^n + i\hem d\btheta_i \swedge d\theta^i$,
so
the nonvanishing Poisson brackets are
$\pb{x^m}{p_n} = \delta^m{}_n$, 
$\pb{\psi^m}{\psi^n} = -i\mem \eta^{mn}$,
and
$\pb{\theta^i}{\btheta_i} = -i\mem \delta^i{}_j$.
The astute reader will notice that
this provides a symplectic realization of 
our earlier Poisson manifold in \eqref{xpq-free},
up to the fermionic extension by the spin variable $\psi^m$.
This means that
the particle's color charge
is recast as a composite variable,
\begin{align}
	\label{qdef}
	q_a \,:=\, i\, \btheta_i\mem (t_a)^i{}_j\mem \theta^j
	\qiq
	\pb{q_a}{q_b}
	\,=\,
		q_c\mem f^c{}_{ab}
	\,,
\end{align}
where $(t_a)^i{}_j$ are the generators
in the representation $\rho$.
Note that the strict requirement for 
the Feynman derivation
is a formulation of a classical particle in a Poisson manifold,
not necessarily symplectic.

Notably, the phase space $\Aflat_\YM$ enjoys
a $\N {\,=\,} 1$ supersymmetry.
The supercharge $Q {\:=\:} p_m \psi^m$
defines the Hamiltonian as $H = \frac{i}{2}\mem \pb{Q}{Q} = \frac{1}{2}\mem p^2$.
By adopting the well-known Dirac framework of constrained Hamiltonian mechanics
\cite{dirac1950generalized,henneaux1992quantization},
we then take
$Q \approx 0$ and $H \approx 0$ as first-class constraints,
so
the resulting
constrained phase space $\Aflat_{\YM*}$
describes a massless on-shell colored spinning particle.
Geometrically, $\Aflat_{\YM*}$ describes
a symplectic quotient embedded in $\Aflat_\YM$,
realizing
an ambitwistor space
with fermionic extensions.

Strictly speaking,
a complexification should be implemented
to attain the precise definition of ambitwistor space
\cite{Mason:2013sva}.
However, let us work in the non-complexified setup,
regarding the scope
of this paper.

\skip
\paragraph{YM Theory from $\N{\,=\,}1$ Ambitwistor String}%
We now review the worldsheet construction of 
Adamo, Casali, and Nekovar \cite{adamo-ym}.
A closed, chiral string theory is given 
for a symplectic target,
featuring the following operator product expansions (OPEs) between
$x^m$, $p_m$, $\psi^m$, and $j_a$
in the free theory:
\begin{subequations}
\label{wsOPE}
\begin{align}
	\label{wsOPE.xp}
	x^m(\s')\, p_n(\s)
	&\,\,\sim\,\,
		x^m p_n
		+
		\frac{
			\delta^m{}_n
		}{\s'{\mem-\,}\s}
	\,,\\
	\label{wsOPE.psi}
	\psi^m(\s')\, \psi^n(\s)
	&\,\,\sim\,\,
		\psi^m \psi^n
		+
		\frac{
			\eta^{mn}
		}{\s'{\mem-\,}\s}
	\,,\\
	\label{wsOPE.j}
	j_a(\s')\, j_b(\s)
	&\,\,\sim\,\,
		j_a j_b
		+
		\frac{
			j_c f^c{}_{ab}
		}{\s'{\mem-\,}\s}
		+ 
		\frac{
			k\mem \delta_{ab}
		}{(\s'{\mem-\,}\s)^2}
	\,,
\end{align}
\end{subequations}
where single contraction yields a single pole $1/(\s'{\mem-\,}\s)$,
double contraction yields a double pole $1/(\s'{\mem-\,}\s)^2$,
and so on.
It is important that the model is chiral,
so the OPEs are meromorphic as such.
$k$ denotes the level of the worldsheet current algebra.

To descend to the ambitwistor space
as a constraint surface,
one imposes
the vanishing of
the supercharge $Q$
and the Hamiltonian $H$
as constraints.
In the classical theory,
we recall that
the Dirac framework \cite{dirac1950generalized,henneaux1992quantization}
has demanded the first-class condition,
meaning 
that
$Q$ and $H$ form a closed algebra under the Poisson bracket.
In the quantum theory,
a natural generalization is to demand that
$Q$ and $H$ form a closed algebra under the OPE,
which might be called the quantum first-class condition.

\rcite{adamo-ym},
specifically,
examines the
$QQ$ and $QH$ OPEs.
When evaluated in curved backgrounds, 
the $QH$ OPE
derives
the magnetic-type and electric-type YM equations
from
the single and double contractions,
respectively.

\paragraph{From Strings to Particles}%
In this paper,
we formulate the worldline counterpart of \rcite{adamo-ym}'s construction
as a sigma model $\R \to \Aflat_\YM$.
As highlighted in the introduction,
this is facilitated by the very chiral nature of the ambitwistor worldsheet model.
For instance,
the worldline action is deduced 
by simply replacing the $\bar{\partial}$ operator in \rcite{adamo-ym}'s worldsheet action
with the worldline time derivative:
$\int (p_m \dot{x}^m {\mem+\,} \frac{i}{2}\mem \eta_{mn}\mem \psi^m \dot{\psi}^n {\mem+\,} i\mem \bar{\theta}_i\mem \dot{\theta}^i)\mem d\t$.
Then, by using the time-symmetric propagator,
we compute
the expectation value of 
two worldline operators
inserted at times $\t' = \t+\e$ and $\t$
with a small $\e>0$.
The result is
\begin{subequations}
\label{wlOPE}
\begin{align}
	\label{wlOPE.xp}
	x^m(\t')\, p_n(\t)
	&\,\,\sim\,\,
		x^m p_n
		+ 
		\tfrac{i\hbar}{2}\,
			\delta^m{}_n
	\,,\\
	\label{wlOPE.xp}
	\psi^m(\t')\, \psi^n(\t)
	&\,\,\sim\,\,
		\psi^m \psi^n
		+ 
		\tfrac{\hbar}{2}\,
			\eta^{mn}
	\,,\\
	\label{wlOPE.q}
	q_a(\t')\, q_b(\t)
	&\,\,\sim\,\,
		q_a q_b
		+
		\tfrac{i\hbar}{2}\,
			q_c f^c{}_{ab}
		+ 
		(\tfrac{i\hbar}{2})^2\,
			k\mem \delta_{ab}
	\,,
\end{align}
\end{subequations}
where 
single contraction yields an $\O(\hbar^1)$ term,
double contraction yields an $\O(\hbar^2)$ term,
and so on.
The double contraction in \eqref{wlOPE.q}
arises because $q_a$ is defined as a composite variable
in \eqref{qdef}.
The constant $k$ is given by
$(t_a)^i{}_j\hem (t_b)^j{}_i = k\mem \delta_{ab}$.

Evidently, the worldline OPEs in \eqref{wlOPE}
precisely mirror the worldsheet OPEs in \eqref{wsOPE},
where the worldline charge $q_a$ corresponds to the worldsheet current $j_a$.

\begin{figure}
	\begin{align*}
		\Big\langle\,\,
		\adjustbox{valign=c}{\begin{tikzpicture}[]
				\node (F) at (0,0) {};
				\node (G) at (0.65,0) {};
				\node (xshift) at (0.2,0) {};
				\node[empty] (Fe) at ($(F)-(xshift)$) {};
				\node[empty] (Ge) at ($(G)+(xshift)$) {};
				\draw[dprop] (Fe)--(F)--(G)--(Ge);
				\node[dot] (f) at ($(F)$) {};
				\node[dot] (g) at ($(G)$) {};
		\end{tikzpicture}}
		\,\,\Big\rangle
		\,\,\,=\,\,\,
		\adjustbox{valign=c}{\begin{tikzpicture}[]
				\node (F) at (0,0) {};
				\node (G) at (0.65,0) {};
				\node (xshift) at (0.2,0) {};
				\node[empty] (Fe) at ($(F)-(xshift)$) {};
				\node[empty] (Ge) at ($(G)+(xshift)$) {};
				\draw[dprop] (Fe)--(F)--(G)--(Ge);
				\node[dot] (f) at ($(F)$) {};
				\node[dot] (g) at ($(G)$) {};
		\end{tikzpicture}}
		\,\,+\,\,
		\adjustbox{valign=c}{\begin{tikzpicture}[]
				\node (F) at (0,0) {};
				\node (G) at (0.65,0) {};
				\node (xshift) at (0.2,0) {};
				\node[empty] (Fe) at ($(F)-(xshift)$) {};
				\node[empty] (Ge) at ($(G)+(xshift)$) {};
				\draw[dprop] (Fe)--(F)--(G)--(Ge);
				\node[dot] (f) at ($(F)$) {};
				\node[dot] (g) at ($(G)$) {};
				\draw[l] (F) -- (G);
		\end{tikzpicture}}
		\,\,+\,\,
		\adjustbox{valign=c}{\begin{tikzpicture}[]
				\node (F) at (0,0) {};
				\node (G) at (0.65,0) {};
				\node (xshift) at (0.2,0) {};
				\node[empty] (Fe) at ($(F)-(xshift)$) {};
				\node[empty] (Ge) at ($(G)+(xshift)$) {};
				\draw[dprop] (Fe)--(F)--(G)--(Ge);
				\node[dot] (f) at ($(F)$) {};
				\node[dot] (g) at ($(G)$) {};
				\draw[l] (F) to[out=+50, in=+130] (G);
				\draw[l] (F) to[out=-50, in=-130] (G);
		\end{tikzpicture}}
		\,\,+\,\,
			\cdots
	\end{align*}
	\caption{
		The Moyal star product from path integral.
	}
	\label{fig:moyal}
\end{figure}

\skip
\paragraph{Worldline OPE as Moyal Star Product}%
As is nicely established in \rcite{deformation-quantification},
the general formula for
such
free worldline OPE
is given by
$\O_1(\t')\mem \O_2(\t) \sim \O_1\hnem \star \O_2$,
where
$\star$ denotes the so-called
Moyal star product \cite{Moyal:1949sk,Groenewold:1946kp}:
\begin{align}
	\label{def-moyal}
	&
	\O_1 \hnem\star \O_2
	\\
	&
	\,=\,
		\O_1 \O_2
		+ \tfrac{i\hbar}{2}\mem
			\pb{\O_1}{\O_2}
		+ \tfrac{1}{2!}\mem (\tfrac{i\hbar}{2})^2\mem
			\dpb{\O_1}{\O_2}
		+ \cdots
	\,.
	\nonumber
\end{align}
Here, we have denoted
\begin{subequations}
\label{br}
\begin{align}
	\label{pb}
	\pb{\O_1}{\O_2}
	&\mem:=\mem
		(\partial_I \O_1)
		\,
				\Pi^{IJ}
		(\partial_J \O_2)
	\,,\\
	\label{dpb}
	\dpb{\O_1}{\O_2}
	&\mem:=\mem
		(\partial_I \partial_K \O_1)
		\,
				\Pi^{IJ}\hem \Pi^{KL}
		\mem
		(\partial_J \partial_L \O_2)
	\,,
\end{align}
\end{subequations}
while $\Pi^{IJ}$ are the (constant) components of the Poisson bivector
in the coordinates $X^I = (x^m, p_m, \psi^m, \theta^i, \btheta_i)$.
Of course,
\eqref{pb} is the very Poisson bracket.
\eqref{dpb}, however,
is a symmetric second-order bi-differential operator
which we dub the \textit{double Poisson bracket}.

This fact is easily derived in the path integral formalism.
Each worldline propagator
describes $i\hbar$ times a Poisson bracket,
together with a step function valued in $\pm \frac{1}{2}$.
The expectation value is diagrammatically computed as in \fref{fig:moyal}
and thus gives rise to the formula in \eqref{def-moyal}.
Especially, the double Poisson bracket 
arises from the one-loop diagram
in \fref{fig:moyal},
computing the double contraction.
Otherwise, one can also note that
\eqref{def-moyal} simply implements Wick's theorem
for 
the Weyl (symmetric) ordering in the operator formalism:
$\hat{x}^m \hat{p}_n = \normal{\hat{x}^m \hat{p}_n} + \frac{i\hbar}{2}\mem \delta^m{}_n$, etc.

With these understandings,
we conclude that
the complete statement of the quantum first-class condition for our ambitwistor particle is
\begin{align}
	\label{qfirstclass}
	Q \star Q \,=\, \hbar\hem H 
	\,,\quad
	Q \star\hnem H \,=\, QH
	\,,\quad
	H \star H \,=\, H^2
	\,.
\end{align}
Up to $\O(\hbar^2)$,
the nontrivial implications of \eqref{qfirstclass}
are
\begin{align}
	\label{master}
	\pb{Q}{H} \mem=\mem 0
	\,,\,\,\,\,
	\dpb{Q}{H} \mem=\mem 0
	\,,\,\,\,\,
	\dpb{H}{H} \mem=\mem 0
	\,.
\end{align}

Finally, by
recalling \rcite{adamo-ym}'s result on the worldsheet,
we expect that
$\pb{Q}{H} {\,=\,} 0$ and $\dpb{Q}{H} {\,=\,} 0$,
evaluating the single and double contractions
in the worldline $QH$ OPE,
will derive
the magnetic-type and electric-type YM equations,
respectively.

In sum,
we have extracted the essence of \rcite{adamo-ym}'s worldsheet construction
and provided a worldline formulation
in \eqref{qfirstclass},
for which 
the Poisson and double Poisson brackets
compute the single and double contractions.

\skip
\paragraph{Covariantized Brackets}%
Strictly speaking, 
however,
the astute reader will point out that
we have established 
the exact correspondence between
the chiral worldsheet OPE
and 
the worldline OPE
only in the free theory limit.
And unfortunately, 
a caveat indeed arises in curved backgrounds.

In \rcite{adamo-ym},
the authors use
the free-theory OPEs in \eqref{wsOPE}
also in curved backgrounds
by utilizing canonical coordinates on the target.
For the worldline,
this means to define
the Moyal star product in \eqref{def-moyal}
with respect to the canonical momentum $p_m^\can$.

We have explicitly checked through brute-force calculations
that such an approach for the worldline
does not yield gauge-covariant equations.
The failures involve various cases of bare gauge potential $A$,
and it seems impossible that 
a single gauge choice can make them vanish altogether.
This issue may be traced to
subtle differences between
the worldsheet and worldline OPEs.

From a geometrical standpoint,
the above failure could be attributed to the fact that
$p^\text{can}_m = p_m + $ $q_a A^a{}_m(x)$
describes a gauge-dependent coordinate transformation 
in phase space
from the chart
$(x^m,p_m,\psi^m,\theta^i,\btheta_i)$
to a Darboux chart
$(x^m,p_m^\text{can},\psi^m,\theta^i,\btheta_i)$.
Crucially, the double Poisson bracket in \eqref{dpb}
is not invariant
under 
coordinate 
change,
since 
second derivatives are not tensors.

By building upon this line of thought,
we have discovered that
a resolution 
is viable
through covariantizing
the brackets in \eqref{br}:
\begin{subequations}
\label{cbr}
\begin{align}
	\label{cpb}
	\pb{\O_1}{\O_2}_\cov
	&\mem:=\mem
		(\nabla_{\nem I} \O_1)\, \Pi^{IJ} (\nabla_{\nem J} \O_2)
	\,,\\
	\label{cdpb}
	\dpb{\O_1}{\O_2}_\cov
	&\mem:=\mem
		(\nabla_{\nem I}\hnem \nabla_{\nem K} \O_1)
		\,
			\Pi^{IJ}\hem \Pi^{KL}
		\mem
		(\nabla_{\nem J}\hnem \nabla_{\nem L} \O_2)
	\,.
\end{align}
\end{subequations}
Here, $\nabla$ is 
a torsion-free affine connection
on the phase space
that preserves the symplectic structure.
It is known that 
such a connection always exists
and is even not unique
\cite{fedosov1994simple,gelfand1997fedosov}.
From physical grounds,
we further stipulate that $\nabla$ is invariant under
the gauge transformations induced in the phase space
due to the spacetime fields.
As a result,
a unique choice seems to stand out for each system,
based on methods we have developed in \rcite{gde}.

To clarify,
\eqref{cpb} simply coincides with \eqref{pb}
as long as $\O_1,\O_2$ are scalars in the phase space.
However, \eqref{cdpb} is distinct from \eqref{dpb}:
the former is tensorial while the latter is not.

\skip
\paragraph{The Master Equations}%
Via this final refinement,
we arrive at the very finding that
the first-order and second-order field equations
of YM theory and gravity
are derived by
the covariantized counterpart of \eqref{master}:
\begin{subequations}
\label{MASTER}
\begin{align}
	\label{1M}
	\pb{Q}{H} \mem=\mem 0
	&\,\,\,\implies\,\,\,
		\textit{first-order,\:magnetic}
	\,,\\
	\label{1E}
	\dpb{Q}{H}_\cov \mem=\mem 0
	&\,\,\,\implies\,\,\,
		\textit{first-order,\:electric}
	\,,\\
	\label{2}
	\dpb{H}{H}_\cov \mem=\mem 0
	&\,\,\,\implies\,\,\,
		\textit{second-order}
	\,.
\end{align}
\end{subequations}
The Hamiltonians are
\smash{$ H \sim \frac{1}{2}\mem p^2 - \frac{1}{2}\mem \btheta \theta F \psi \psi $}
for YM and
\smash{$ H \sim \frac{1}{2}\mem p^2 - \frac{1}{2}\mem \bpsi \psi R \bpsi \psi $}
for gravity.
A detailed verification is provided in the appendices.
Below, we briefly summarize the results.

\skip
\paragraph{YM Theory from $\N{\,=\,}1$ Ambitwistor Particle}%
For YM theory, 
we take the curved version of 
the space $\Aflat_\YM$ in \eqref{Aflat-YM}
in terms of the Souriau-Feynman \cite{souriau1970structure,dyson1990feynman} deformation of the symplectic structure:
\begin{align}
	\label{A-YM}
	\A_\YM
	\,=\,
		( T^* {\hem\oplus\mem} \Pi T ) \mflat
		\,\oplus E
	\,.
\end{align}
Here, 
$E$ is a vector bundle over $\mflat$ whose typical fiber is $T^*\Pi\V$.
The symplectic form is
$\omega = d( p_m dx^m {\,+\,} \frac{i}{2}\mem \eta_{mn} \psi^m $
$d\psi^n {\,+\,} i\hem \btheta_i\hem D\theta^i )$,
where $D\theta^i = d\theta^i + (t_a)^i{}_j\mem \theta^j\mem A^a{}_m(x)\mem dx^m$.
With an explicit construction of a gauge-invariant symplectic torsion-free connection $\nabla$,
we evaluate \eqref{MASTER} in $\A_\YM$.
This derives
not only
\eqrefs{YM.1m}{YM.1e}
but also 
the second-order equations on the YM field strength
put forward by Cheung and Mangan \cite{cck}
for a covariant cousin of color-kinematics duality:
\begin{align}
	\label{YM.2}
	D^2 F^a{}_{mn} + 2\mem f^a{}_{bc}\mem F^b{}_{mr} F^{cr}{}_n = 0
	\,.
\end{align}
In this process,
we take the formal limit of $k {\,\to\;} 0$
like in \rcite{adamo-ym}.
Note that $k$ precisely arises due to the compositeness of the color charge $q_a$ in the symplectic realization,
whereas the physical essence of the Feynman logic
has required only Poisson manifolds.

In the above derivation,
the key parts of 
\eqrefs{1M}{1E}
arise as
$
	\pb{\pb{p_m}{p_n}}{p_r}\mem \psi^m \psi^n  \psi^r
	{\,\sim\,} \pb{\pb{Q}{Q}}{Q}
$
and 

\noindent
$\{\{p_m,p_n\},p^n\} {\hem\sim\hem} \{\pb{p_m}{p_n},p_r\}\mem \pb{\psi^n}{\psi^r} {\hem\sim\hem} \dpb{\pb{Q}{Q}}{Q}$.
In this manner, \eqref{!magnetic} reincarnates as a consistency for supersymmetry
while \eqref{!electric} emerges through a fermionic contraction.
Amusingly,
the worldline fermion $\psi^m$
universally implements
both
the index antisymmetrization for magnetic-type equations
and the index contraction for electric-type equations.

\skip
\paragraph{General Relativity from $\N{\,=\,}2$ Ambitwistor Particle}%
For general relativity, we use the ambitwistor space construction
with $\N{\,=\,}2$ supersymmetry,
following Adamo, Casali, and Skinner \cite{adamo-gravity}:
\begin{align}
	\label{A-Grav}
	\A_\Grav
	\,=\,
		( T^* {\hem\oplus\mem} \Pi T^\C ) \M
	\,.
\end{align}
Here, $\M {\,=\,} (\R^d,g)$ is a real Riemannian manifold.
The symplectic form is
$\omega = d( p_m e^m {\,+\,}  
i\hem \bpsi_m D\psi^m)$,
where 
$e^m$ is the orthonormal coframe,
and
$D$ encodes the spin connection.
With an explicit gauge-invariant symplectic torsion-free connection $\nabla$,
we evaluate \eqref{MASTER} in $\A_\Grav$
for one of the supercharges, $Q = p_m\psi^m$.
The output is Ricci flatness,
the magnetic and electric equations \cite{cck}
\begin{align}
	D_\wrap{[k} R^{mn}{}_\wrap{rs]}
	\,=\,
		0
	\,,\quad
	D^r R^{mn}{}_{rs}
	\,=\, 
		0
	\,,
\end{align}
and the second-order equations on the Riemann tensor
known as Penrose wave equation
\cite{Penrose:1960eq,Ryan:1974nt}:
\begin{align}
\left({\,
\begin{aligned}
	&
	D^2 R^m{}_n{}^r{}_s
	- R^m{}_n{}^k{}_l R^l{}_k{}^r{}_s
	\\
	&
	+ 2\mem R^m{}_k{}^r{}_l R^k{}_n{}^l{}_s 
 	- 2\mem R^m{}_k{}^l{}_s R^k{}_n{}^r{}_l
\end{aligned}
}\,\right)
	\,=\,
		 0
	\,.
\end{align}
Further, we have checked that
the Kalb-Ramond field $B_{\m\n}$
can be incorporated
by deforming the supercharges as
$Q = p_m \psi^m - \frac{i}{2}\mem (D_\wrap{r} B_\wrap{mn})\mem \psi^r \psi^m \psi^n + (D_k D_\wrap{r} B_\wrap{mn}) \psi^k \psi^r $ $ \psi^m \psi^n$,
just like in \rcite{adamo-gravity}.

\paragraph{Gauge Covariance Versus Associativity: Quantum}%
Finally, we want to interpret \eqref{MASTER}
as the quantum first-class condition
for the gauge-covariant constrained quantization of curved ambitwistor spaces $\A_\YM, \A_\Grav$:
\begin{align}
	\label{qfirstclass-cov}
	Q \star_\cov \nem\nem Q \mem=\mem \hbar\hem H 
	\,,\,\,\,\,
	Q \star_\cov\hnem \nem\nem H \mem=\mem QH
	\,,\,\,\,\,
	H \hnem \star_\cov \nem\nem H \mem=\mem H^2
	\,.
\end{align}
This supposes the existence of gauge-covariant and, crucially, associative operator algebras $\star_\nabla$
such that
\begin{align}
	\label{def-fed}
	&
	\O_1 \hnem\star_\cov \O_2
	\\
	&
	\,=\,
		\O_1 \O_2
		+ \tfrac{i\hbar}{2}\mem
			\pb{\O_1}{\O_2}_\cov
		+ \tfrac{1}{2!}\mem (\tfrac{i\hbar}{2})^2\mem
			\dpb{\O_1}{\O_2}_\cov
		+ \cdots
	\,.
	\nonumber
\end{align}
The proof of constructive existence of such algebras
is immediate by the Fedosov \cite{fedosov1994simple} theory.
For each symplectic torsion-free connection $\nabla$,
Fedosov constructs a unique associative star product
whose expansion is given by \eqref{def-fed}
\footnote{
	The uniqueness holds under natural assumptions.
	Also, 
	in \oldeqref{def-fed}, we have hidden an $\O(\hbar^2)$ term inside the ellipsis:
	$
		- \tfrac{2}{3}\mem (\tfrac{i\hbar}{2})^2\,
			\Ric[\nabla]^{IJ}
				(\nabla_I \O_1)\mem
				(\nabla_J \O_2)
	$.
	Amusingly, there is no caveat in our statement that
	\oldeqref{qfirstclass-cov} up to $\O(\hbar^2)$ implies \oldeqref{MASTER}
	because
	the Ricci tensor of our $\nabla$ vanishes
	(identically for YM, on spacetime Ricci flatness for gravity).
}.
Since our connections $\nabla$ are gauge invariant,
the Fedosov framework
defines gauge-covariant associative star products
on the ambitwistor worldlines.

The way how Fedosov reconciles gauge covariance and associativity
is interesting:
a fiberwise Moyal star product is employed
while
the curved linear connection $\nabla$ is deformed into 
a flat nonlinear connection,
reminscently of 
curved twistor theory
\cite{Penrose:1976js,Mason:2009afn}.

\newpage

\eqref{def-fed} 
could also be approached
by performing the path integral
in normal coordinates due to $\nabla$,
in which case 
\eqref{cdpb}
evaluates a one-loop diagram.

\skip
\paragraph{BAS Formulation of Gravity}%
This last step of promoting \eqref{MASTER} to \eqref{qfirstclass-cov}
is crucial
for our conjecture.
If the Fedosov star product
for gravity
can somehow be brought to 
the Moyal star product
in canonical coordinates
via certain methods
\cite{kontsevich,cattaneo-felder,fedosov1994simple},
one may formulate gravity by
the BAS equations in \eqref{conjecture}
based on a strictly nondynamical double Poisson bracket.
As established in \rcite{cck},
a theory exhibits
BCJ duality
at tree level
if its equations of motion can be formulated as
the BAS equations
for some choice of the Lie algebras.
Thus, one manifests tree-level BCJ duality for gravity.

\skip
\paragraph{Conclusions}%
In this paper,
we provided a unified formulation of YM theory and gravity
from ambitwistor spaces.
%
%
Physically, this imposes
consistency of gauge-covariant constrained quantization,
from which
double Poisson brackets
arise as one-loop diagrams.

Ambitwistor space has been a hope for
understanding gauge theory and gravity
without self-duality restrictions
and
in general dimensions
\cite{Witten:1978ambi,Isenberg:1978kk,Geyer:2022cey}.
It will be exciting if the elusive kinematic algebra 
manifests
via
the grammar of double Poisson bracket
for ambitwistor space,
generalizing the heavenly equation
for self-dual gravity.

A closed string picture for our construction 
might be viable,
on account of established results \cite{kontsevich,cattaneo-felder}.

	\medskip
	\noindent\textit{Acknowledgements.}|%
		The author is grateful to
			Clifford Cheung,
			Toby Saunders-A'Court,
		and
			Sonja Klisch
		for discussions.
		The author would like to thank
			Lionel Mason
		for insightful discussions
		during the
		conference ``The Mathematics behind Scattering Amplitudes'' held in August 2024;
		the author thanks the Galileo Galilei
		Institute for Theoretical Physics, Florence for hospitality. 
		The author is grateful to
			Sebastian Mizera
		for encouraging comments
		and
			Julio Parra-Martinez
		for bringing the history of Feynman brackets to his attention.
		The author thanks
		the attendees of California Amplitudes Meeting 2023 on March 18\textsuperscript{th} for 
		comments on the idea of approaching double copy via Feynman brackets,
		presented in the talk \cite{caamps23spt}.
		J.-H.K. is supported by the Department of Energy (Grant {No.}~DE-SC0011632) and by the Walter Burke Institute for Theoretical Physics.

	\newpage
	\appendix
	\onecolumngrid
	\section{Covariant Symplectic Geometry for YM Theory}
\label{APP.YM}

\skip
\para{Frame}%
The geometry of the space $\A_\YM$ in \eqref{A-YM}
admits a manifestly gauge-covariant description
in terms of 
the noncanonical coordinates $(x^m,p_m,\psi^m,\theta^i,\btheta_i)$
and a noncoordianate frame
$\E_A = (\X_r,\P^m,\Ps_m,\Th_i,\bTh^i)$:
\begin{align}
	\label{YM.frame}
	\X_r
	\mem&=\mem
		\frac{\partial}{\partial x^r}
		- 
			\bigbig{\LTh A_r\theta}
		+ 
			\bigbig{\btheta A_r \RbTh}
	\,,\quad
	\P^m
	\mem=\mem
		\frac{\partial}{\partial p_m}
	\,,\quad
	\Ps_m
	\mem=\mem
		\frac{\partial}{\partial \psi^m}
	\,,\quad
	\Th_i
	\mem=\mem
		\frac{\partial}{\partial \theta^i}
	\,,\quad
	\bTh^i
	\mem=\mem
		\frac{\partial}{\partial \btheta_i}
	\,.
\end{align}
Here, 
we have abbreviated contracted indices as
$\bigbig{\LTh A_r\theta} = \LTh_i\hem A^i{}_{jr}\mem \theta^j$,
where $A^i{}_{jr} := (t_a)^i{}_j\hem A^a{}_r$.
The accents 
$\footnotesize\blacktriangleleft$
and
$\footnotesize\blacktriangleright$
specify directionalities for fermionic derivatives.
The gauge covariance of this frame
is immediate by the fact that
it is dual to the gauge-covariant basis of one-forms,
$(dx^m , dp_m , d\psi^m , D\theta^i , D\btheta_i)$.
In particular, 
$\X_r$ 
is the very horizontal lift \cite{ehresmann1948connexions,Mason:2013sva}
of the spacetime derivative by the nonabelian gauge connection.
The computation of the Lie brackets $[\E_A,\E_B]$
is left as an exercise.

\skip
\para{Symplectic Connection}%
In the above gauge-covariant noncoordinate basis, the Poisson bivector of $\A_\YM$ reads
\begin{align}
	\label{ym.pi}
	\Pi
	\,=\,
		\X_m \mwedge \P^m
		- i\hem \eta_{mn}\mem
			\tfrac{1}{2}\mem \Ps_m \mwedge \Ps_n
		- i\mem \Th_i \mwedge \bTh^i
		+ \tfrac{1}{2}\mem 
			\bigbig{ i\hem \btheta\hem F_{mn} \hem\theta }\mem \P^m \swedge \P^n
	\,.
\end{align}
To construct the phase space connection $\nabla$ in $\A_\YM$,
we impose
the Poisson-preserving condition
$\nabla_{\nem\E_A} \Pi = 0$
as well as
the torsion-free condition 
$\nabla_{\nem\E_A}\hnem \E_B - \nabla_{\nem\E_B}\hnem \E_A = [\E_A,\E_B]$.
For the former,
$\E_A( i\hem \btheta\mem F_{mn}\mem \theta ) \neq 0$
implies
some necessary cancellations
via nonzero connection coefficients,
which, in turn, propagate to other components through the torsion-free condition.
Iterating, 
we find that the following choice for 
the nonvanishing connection coefficients 
defines a possible instance of a torsion-free symplectic connection,
provided the Bianchi identity $D_\wrap{[m} F^a{}_\wrap{nk]} = 0$:
\begin{subequations}
\label{ym.cc}
\begin{align}
	\nabla_{\nem\X_m} \X_n
	\,&=\,
		\hlg{{
			- \tfrac{1}{2}\mem \bigbig{ \Th F_{mn} \theta }
			+ \tfrac{1}{2}\mem \bigbig{ \btheta F_{mn} \bTh }
		}}
		\hld{
			+ \tfrac{2}{3}\mem 
				\bigbig{ i\hem \btheta\hem D_\wrap{(m} F_\wrap{n)k} \hem\theta }
				\mem \P^k
		}
	\,,\\
	\nabla_{\nem\X_r} \LTh_i
	\,&=\,
		\hlc{{
			+ ( \LTh A_r )_i
		}}
		\hlg{
			+ \tfrac{1}{2}\mem (i\hem \btheta F_{rk})_i\mem \P^k
		}
	\,,\quad
	\nabla_{\nem\la{\Th}_i} \X_r
	\,=\,
		\hlg{{
			+ \tfrac{1}{2}\mem (i\hem \btheta F_{rk})_i\mem \P^k
		}}
	\,,\\
	\nabla_{\nem\X_r} \RbTh^i
	\,&=\,
		\hlc{{
			- ( A_r \RbTh )^i
		}}
		\hlg{
			+ \tfrac{1}{2}\mem (i F_{rk} \theta)^i\mem \P^k
		}
	\,,\quad
	\nabla_{\nem\ra{\bTh}^i} \X_r
	\,=\,
		\hlg{{
			+ \tfrac{1}{2}\mem (i F_{rk} \theta)^i\mem \P^k
		}}
	\,.
\end{align}
\end{subequations}
The gauge invariance of this $\nabla$
is explicitly checked
by using the gauge-covariant transformation behavior of $\E_A$.

\skip
\para{Covariant Hessian}%
Let
$\nabla^2\O := (\nabla_{\nem I}\nabla_{\nem J}\O)\mem \partial_I {\mem\otimes\mem} \partial_J$
be the covariant Hessian of a scalar $\O$,
which is a symmetric tensor.
In the noncoordinate basis $\E_A$,
its components 
can be computed as
$(\nabla^2\O)(\E_A,\E_B) = \E_A(\E_B\O) - (\nabla_{\E_A}\hnem\E_B)\hem \O$.

In the space $\A_\YM$,
the supercharge $Q$ and the Hamiltonian $H$ are given by
\begin{align}
	Q \mem=\mem p_m \psi^m
	\,,\quad
	H \mem=\mem H_0 + H_1
	\,,\quad
		H_0 \mem=\mem \tfrac{1}{2}\mem p^2
	\,,\quad
		H_1 \mem=\mem 
			-\tfrac{1}{2}\mem 
				\btheta_i\mem \theta^j F^i{}_{jmn}\mem \psi^m \psi^n
	\,.
\end{align}
Computation shows that
the nonvanishing components of
their covariant Hessians
are
\begin{subequations}
\begin{align}
\begin{split}
\label{ym.hess-Q}
	(\nabla^2Q)(
		\X_m , \LTh_i
	)
	\mem=\mem
		+ \tfrac{1}{2}\mem (i\hem \btheta F_{mk})_i\mem \psi^k
&\,,\quad
	(\nabla^2Q)(
		\P^m , \Ps_n
	)
	\mem=\mem
		\delta^m{}_n
\,,\\
	(\nabla^2Q)(
		\X_m , \RbTh^i
	)
	\mem=\mem
		- \tfrac{1}{2}\mem (i\hem F_{mk} \theta)^i\mem \psi^k
&\,,\quad
	(\nabla^2Q)(
		\X_m , \X_n
	)
	\mem=\mem
		- \tfrac{2}{3}\mem 
			\bigbig{ i\hem \btheta\mem D_\wrap{(m} F_\wrap{n)k} \mem\theta }
			\mem \psi^k
	\,,
\end{split}\\[0.25\baselineskip]
\begin{split}
\label{ym.hess-0}
	(\nabla^2H_0)(
		\X_r , \LTh_i
	)
	\mem=\mem
		-\tfrac{1}{2}\mem 
			(i\hem \btheta F_{rk})_i\mem p^k
&\,,\quad
	(\nabla^2H_0)(
		\P^m , \P^n
	)
	\mem=\mem
		\eta^{mn}
\,,\\
	(\nabla^2H_0)(
		\X_r , \RbTh^i
	)
	\mem=\mem
		-\tfrac{1}{2}\mem 
			(i\hem F_{rk} \theta)^i\hem p^k
&\,,\quad
	(\nabla^2H_0)(
		\X_m , \X_n
	)
	\mem=\mem
		- \tfrac{2}{3}\mem 
			\bigbig{ i\hem \btheta\mem D_\wrap{(m} F_\wrap{n)k} \mem\theta }
			\mem p^k	
	\,,
\end{split}\\[0.25\baselineskip]
\begin{split}
\label{ym.hess-1}
	(\nabla^2 H_1 )(
		\X_m , \LTh_i
	)
	\mem=\mem
		- \tfrac{1}{2}\mem (\btheta\hem D_m F_{rs})_i\mem \psi^r \psi^s
&\,,\quad
	(\nabla^2 H_1 )(
		\X_m , \X_n
	)
	\mem=\mem
		- \tfrac{1}{2}\mem 
		\bigbig{ \btheta\hem D_\wrap{(m} D_\wrap{n)} F_{rs} \theta }\mem \psi^r \psi^s
\,,\\
	(\nabla^2 H_1 )(
		\X_m , \RbTh^i
	)
	\mem=\mem
		- \tfrac{1}{2}\mem (D_m F_{rs} \theta)^i\mem \psi^r \psi^s
&\,,\quad
	(\nabla^2 H_1 )(
		\RbTh^i , \LTh_j
	)
	\mem=\mem
		- \tfrac{1}{2}\mem F^i{}_{jrs}\mem \psi^r \psi^s
\,,
\end{split}\\[0.25\baselineskip]
\begin{split}
	(\nabla^2 H_1 )(
		\LPs_r , \LTh_i
	)
	\mem=\mem
		- (\btheta F_{rs})_i\mem \psi^s
&\,,\quad
	(\nabla^2 H_1 )(
		\X_m , \RA{\Ps}_r
	)
	\mem=\mem
		- \bigbig{\btheta D_m F_{rs} \theta}\mem \psi^s
\,,\\
	(\nabla^2 H_1 )(
		\RPs_r , \RbTh^i
	)
	\mem=\mem
		+ (F_{rs} \theta)^i\mem \psi^s
&\,,\quad
	(\nabla^2 H_1 )(
		\RA{\Ps}_r , \RA{\Ps}_s
	)
	\mem=\mem
		\bigbig{\btheta\hem F_{rs}\hem \theta}
	\,.
\end{split}
\end{align}
\end{subequations}

\para{Covariant Double Poisson Bracket}%
Now the covariant double Poisson brackets
can be readily computed.
We adopt the consistent, physical convention that
the derivatives in $\dpb{\O_1}{\O_2}_\cov$
act from the right on $\O_1$
and
act from the left on $\O_2$.
The components \smash{$\Pi^{AB}$} 
(where
$ \Pi = \tfrac{1}{2}\mem \Pi^{AB}\mem  \E_A \swedge \E_B$)
are bosonic,
so their ordering 
is irrelevant.

The results,
which are
double-checked by a Mathematica code
based on the xAct and xTerior packages,
are
\begin{subequations}
\begin{align}
	\dpb{Q}{Q}_\cov
	\,&=\,
		0
	\,,\quad
	\dpb{Q}{H}_\cov
	\,=\, 
		- \tfrac{8}{3}\mem 
			\bigbig{ i\mem \btheta\hem D^m F_\wrap{mk} \hem\theta }
			\mem \psi^k
\,,\\
	\dpb{H}{H}_\cov
	\,&=\,
		- \tfrac{4}{3}\mem 
			\bigbig{ i\mem \btheta\hem D^m F_{mk} \theta }
			\mem p^k
		-
			\btheta\mem\BB{
				D^2 F_{mn}
				+ 2\mem [ F_{mr} , F^r{}_n ]
			}\theta
			\, \psi^m \psi^n
		- \tfrac{1}{2}\mem
			k\mem \delta_{ab}\mem
			F^a{}_{mn} F^b{}_{rs}
			\mem
			\psi^m \psi^n \psi^r \psi^s
	\,.
\end{align}
\end{subequations}
The details are shown below:
\begin{subequations}
\begin{align}
	\dpb{Q}{Q}_\cov
	\,&=\,
		\bigbig{ i\hem \btheta F_{mn} \hem\theta }\mem (-i\hem\eta^{mn})
	\contxd{
		\P }{ \P
	}{
		\Ps }{ \Ps
	}\,,\\
	\,&=\, 0
	\,,\\
	\dpb{Q}{H}_\cov
	\,&=\,
	\begin{aligned}[t]
		&{}
		-i\mem \bigbig{ \btheta D^m\hnem F_{mk} \hem\theta }\mem \psi^k
	\contx{
		\P }{ \X
	}{
		\Ps }{ \Ps
	}\\
		&{}
		- \frac{2}{3}\mem 
			\bigbig{ i\hem \btheta D^m F_\wrap{mk} \hem\theta }
			\mem \psi^k
	\contd{
		\X }{ \P
	}{
		\X }{ \P
	}\,,
	\end{aligned}\\
	\label{ym.H0~H0}
	\dpb{H_0}{H_0}_\cov
	\,&=\,
	\begin{aligned}[t]
		&{}
			- \frac{2}{3}\mem 
				\bigbig{ i\hem \btheta D^m F_{mk} \hem\theta }
				\mem p^k
		\conty{
			\X }{ \P
		}{
			\X }{ \P
		}\\
		&{}
			+
				\bigbig{q_a F^{amn}}
				\bigbig{q_b F^b{}_{mn}}
		\contd{
			\P }{ \P
		}{
			\P }{ \P
		}\,,
	\end{aligned}
	\\
	\label{ym.H0~H1}
	\dpb{H_0}{H_1}_\cov
	\,&=\,
	\begin{aligned}[t]
		&{}
			- \frac{1}{2}\mem
				\bigbig{\btheta
					D^2 F_{mn}
				\theta}
				\mem \psi^m \psi^n
		\contd{
			\P }{ \X
		}{
			\P }{ \X
		}\,,
	\end{aligned}
	\\
	\label{ym.H1~H1}
	\dpb{H_1}{H_1}_\cov
	\,&=\,
	\begin{aligned}[t]
		&{}
			- \frac{1}{4}\mem
				k\mem \delta_{ab}\mem
				F^a{}_{mn} F^b{}_{rs}
				\mem
				\psi^m \psi^n \psi^r \psi^s
		\contx{
			\Th }{ \bTh
		}{
			\bTh }{ \Th
		}\\
		&{}
			- \bigbig{\btheta
				F_{mr} F^r{}_n
			\hem\theta}\mem
			\psi^m \psi^n
		\contxy{
			\Th }{ \bTh
		}{
			\Ps }{ \Ps
		}\\
		&{}
			- 
			\bigbig{q_a F^{amn}}
			\bigbig{q_b F^b{}_{mn}}
		\contd{
			\Ps }{ \Ps
		}{
			\Ps }{ \Ps
		}\,.
	\end{aligned}
\end{align}
\end{subequations}
Note that
the last fermionic contraction in \eqref{ym.H1~H1}
cancels the last bosonic contraction in \eqref{ym.H0~H0}
exactly,
for which
the sign factors work out as
\begin{align}
\begin{split}
\wick[offset=1.25em]{
	\Big[\,
		\tfrac{i}{2}\mem
			\bigbig{q_a F^a{}_{rs}}
			\mem \c2 \psi^r \c1 \psi^s
	\,\Big]\Big[\,
		\tfrac{i}{2}\mem
			\bigbig{q_b F^b{}_{kl}}
			\mem \c1 \psi^k \c2 \psi^l
	\,\Big]
}
	\times 2^2
	\,=\,
		(\tfrac{i}{2})^{2}
		\mem
			\bigbig{q_a F^{anm}}
			\mem(-i)^2\mem
			\bigbig{q_b F^b{}_{mn}}
		\times 2^2
	\,=\,
		-
			\bigbig{q_a F^{amn}}
			\bigbig{q_b F_{bmn}}
	\,.
\end{split}
\end{align}

Lastly, the curvature of the phase space connection
is obtained via
$[\nabla_{\nem\E_C},\nem \nabla_{\nem\E_D}]\mem \E_B - \nabla_{\nem[\E_C,\E_D]} \E_B = R^A{}_{BCD}[\nabla]\mem \E_A$.
Brute-force calculation
shows that $R^A{}_{BCD}[\nabla] \neq 0$,
but the Ricci tensor,
$\Ric[\nabla]_{BD} = R^A{}_{BAD}[\nabla]$,
vanishes identically.

\section{Covariant Poisson Geometry for Gravity}
\label{APP.GRAV}

\para{Frame}%
The geometry of the space $\A_\Grav$ in \eqref{A-Grav}
admits a manifestly gauge-covariant description
in terms of 
the noncanonical coordinates $(x^m,p_m,\psi^m,\bpsi_m)$
and a noncoordianate frame
$\E_A = (\X_r,\P^m,\Ps_m,\bPs^m)$:
\begin{align}	
	\label{grav.frame}
	\X_r
	\mem&=\mem
		E^\r{}_r\mem
		\frac{\partial}{\partial x^\r}
		- 	
			\bigbig{\LPs \gamma_r \psi}
		+ 
			\bigbig{\bpsi \gamma_r\nem \RbPs}
		+ 
			\bigbig{p\gamma_r \P}
	\,,\quad
	\P^m
	\mem=\mem
		\frac{\partial}{\partial p_m}
	\,,\quad
	\Ps_m
	\mem=\mem
		\frac{\partial}{\partial \psi^m}
	\,,\quad
	\bPs^m
	\mem=\mem
		\frac{\partial}{\partial \bpsi_m}
	\,.
\end{align}
Here, $E^\m{}_m(x)$ is the vielbein
while $\gamma^m{}_{nr}(x)$ are the spin connection coefficients.
$m,n,\cdots$ are local Lorentz indices
while $\m,\n,\cdots$ are spacetime indices.
The covariance of this frame is immediate by the fact that
it is dual to the covariant basis of one-forms
$(e^m,Dp_m,D\psi^m,D\bpsi_m)$,
where $e^m = e^m{}_\m(x)\mem dx^\m$ is the one-form
dual to $E_m = E^\m{}_m(x)\mem \partial_\m$
while $D$ denotes the Lorentz-covariant derivative.
In particular,
$\X_r$ is the horizontal lift \cite{ehresmann1948connexions,Mason:2013sva} of the spacetime derivative
by the Levi-Civita connection.
It is left as an exercise to compute the Lie brackets $[\E_A,\E_B]$.

\skip
\para{Symplectic Connection}%
In the above gauge-covariant noncoordinate basis, the Poisson bivector of $\A_\Grav$ reads
\begin{align}
	\label{grav.pi}
	\Pi
	\,=\,
		\X_m \mwedge \P^m
		- i\hem \eta_{mn}\mem
			\tfrac{1}{2}\mem \Ps_m \mwedge \Ps_n
		- i\mem \Th_i \mwedge \bTh^i
		+ \tfrac{1}{2}\mem 
			\bigbig{ i\hem \btheta\hem F_{mn} \hem\theta }\mem \P^m \swedge \P^n
	\,.
\end{align}
Taking the same approach as in \App{APP.YM},
we find that the following choice for 
the nonvanishing connection coefficients 
defines a torsion-free symplectic connection
in $\A_\Grav$,
provided the Bianchi identities
$R^m{}_\wrap{[krs]} = 0$,
$D_\wrap{[k} R^m{}_n{}_\wrap{rs]} = 0$
of the Riemann tensor:
\begin{subequations}
\label{grav.ConnectionCoefficients}
\begin{align}
	\nabla_{\nem\X_m} \X_n
	\,&=\,
		\hlc{{
			\gamma^k{}_{nm}\mem \X_k
		}}
		\hlg{
			- \tfrac{1}{2}\mem \bigbig{ \Ps R_{mn} \psi }
			+ \tfrac{1}{2}\mem \bigbig{ \bpsi R_{mn} \bPs }
		}
		\hlx{
			- \tfrac{2}{3}\mem 
				p_l R^l{}_\wrap{(kn)m}\mem \P^k
		}
		\hld{
			+ \tfrac{2}{3}\mem 
				\bigbig{ i\hem \bpsi D_\wrap{(m} R_\wrap{n)k} \psi }
				\mem \P^k
		}
	\,,\\
	\nabla_{\nem\X_r} \LPs_m
	\,&=\,
		\hlc{{
			+ ( \LPs \gamma_r )_m
		}}
		\hlg{
			+ \tfrac{1}{2}\mem (i\hem \bpsi R_{rk})_m\mem \P^k
		}
	\,,\quad
	\nabla_{\nem\lPs_m} \X_r
	\,=\,
		\hlg{{
			+ \tfrac{1}{2}\mem (i\hem \bpsi R_{rk})_m\mem \P^k
		}}
	\,,\\
	\nabla_{\nem\X_r} \RbPs^m
	\,&=\,
		\hlc{{
			- ( \gamma_r \RbPs )^m
		}}
		\hlg{
			+ \tfrac{1}{2}\mem (i\hem R_{rk}\psi)^m\mem \P^k
		}
	\,,\quad
	\nabla_{\nem\rbPs^m} \X_r
	\,=\,
		\hlg{{
			+ \tfrac{1}{2}\mem (i\hem R_{rk}\psi)^m\mem \P^k
		}}
	\,,\\
	\nabla_{\nem\X_r} \P^m
	\,&=\,
		\hlc{{
			- (\gamma_r \P)^m
		}}
	\,.
\end{align}
\end{subequations}
Again, the gauge invariance of this $\nabla$
is explicitly checked
by using the covariant transformation behavior of $\E_A$.

\newpage
\para{Covariant Hessian}%
In the space $\A_\Grav$,
the supersymmetry generators $Q,\bQ$ and the Hamiltonian $H$ 
are given by
\begin{align}
	Q = p_m \psi^m
	\,,\,\,\,
	\bQ = p_m \bpsi^m
	\,,\,\,\,
	H = \tfrac{i}{2}\mem \pb{Q}{\bQ} = H_0 + H_1
	\,,\,\,\,
		H_0 = \tfrac{1}{2}\mem p^2
	\,,\,\,\,
		H_1 
		    = -\tfrac{1}{2}\mem \bpsi_m \psi^n R^m{}_n{}^r{}_s\mem \bpsi_r \psi^s
	\,.
\end{align}
Computation shows that the nonvanishing components of their covariant Hessians in the noncoordinate basis are
\begin{subequations}
\begin{align}
\begin{split}
\label{grav.hess-Q}
	(\nabla^2 Q)(
		\X_r , \LPs_m
	)
	\mem=\mem
		+ \tfrac{1}{2}\mem  
			(i\hem \bpsi R_{rk})_m
			\mem \psi^k
&\,,\quad
	(\nabla^2Q)(
		\P^m , \Ps_n
	)
	\mem=\mem
		\delta^m{}_n
\,,\\
	(\nabla^2 Q)(
		\X_r , \RbPs^m
	)
	\mem=\mem
		- \tfrac{1}{2}\mem 
			(i\hem R_{rk} \psi)^m
			\mem \psi^k
&\,,\quad
	(\nabla^2 Q)(
		\X_r , \X_s
	)
	\mem=\mem
		- \tfrac{2}{3}\mem
			\bigbig{ i\hem \bpsi \nabla_\wrap{\nem(r} R_\wrap{s)k} \psi }\mem
			\psi^k
		- \tfrac{1}{3}\mem 
			p_m
			R^m{}_\wrap{(rs)k}\mem
			\psi^k
	\,,
\end{split}\\[0.25\baselineskip]
\begin{split}
\label{grav.hess-0}
	(\nabla^2H_0)(
		\X_r , \LPs_m
	)
	\mem=\mem
		-\tfrac{1}{2}\mem 
			(i\hem \bpsi R_{rk})_m\mem p^k
&\,,\quad
	(\nabla^2H_0)(
		\P^m , \P^n
	)
	\mem=\mem
		\eta^{mn}
\,,\\
	(\nabla^2H_0)(
		\X_r , \RbPs^m
	)
	\mem=\mem
		-\tfrac{1}{2}\mem 
			(i R_{rk}\psi)^m\mem p^k
&\,,\quad
	(\nabla^2H_0)(
		\X_r , \X_s
	)
	\mem=\mem
		- \tfrac{2}{3}\mem 
			\bigbig{ i\bpsi D_\wrap{(r} R_\wrap{s)k} \psi }
			\mem p^k
		+ \tfrac{1}{3}\mem
			p_m p_n
			R^m{}_r{}^n{}_s	
	\,,
\end{split}\\[0.25\baselineskip]
\begin{split}
\label{grav.hess-1}
	(\nabla^2 H_1 )(
		\X_k , \LPs_m
	)
	\mem=\mem
		- (\bpsi D_k R^r{}_s)_m\mem \bpsi_r \psi^s
&\,,\quad
	(\nabla^2 H_1 )(
		\X_m , \X_n
	)
	\mem=\mem
		-\tfrac{1}{2}\mem 
		\bigbig{ \bpsi D_\wrap{(m} D_\wrap{n)} R^r{}_s \psi }\mem 
			\bpsi_r \psi^s
\,,\\
	(\nabla^2 H_1 )(
		\X_k , \RbPs^m
	)
	\mem=\mem
		- (D_k R^r{}_s \psi)^m\mem \bpsi_r \psi^s
&
\,,
\end{split}\\[0.25\baselineskip]
\begin{split}
	(\nabla^2 H_1 )(
		\LPs_m , \LPs_n
	)
	\mem=\mem
		+
		\bpsi_r \bpsi_s\mem
		R^r{}_m{}^s{}_n
&
\,,\quad
	(\nabla^2 H_1 )(
		\RbPs^m , \LPs_n
	)
	\mem=\mem
		-
		\bpsi_i\mem
		R^{mi}{}_{nj}\mem
		\psi^j
\,,\\
	(\nabla^2 H_1 )(
		\RbPs^m , \RbPs^n
	)
	\mem=\mem
		- R^m{}_r{}^n{}_s\mem 
		\psi^r \psi^s
&
\,.
\end{split}
\end{align}
\end{subequations}
Here,
we have made many uses of 
the Riemann index symmetries.
Observe the parallels between 
\eqrefs{ym.hess-Q}{grav.hess-Q},
\eqrefs{ym.hess-0}{grav.hess-0},
and
\eqrefs{ym.hess-1}{grav.hess-1}:
the fermions $\psi^m$ and $\bpsi_m$ 
serve as color charges.

\skip
\para{Covariant Double Poisson Bracket}%
The covariant double Poisson brackets
are now readily computed.
The results,
which are
double-checked by the Mathematica code mentioned earlier,
read
\begin{subequations}
\begin{align}
	\dpb{Q}{Q}_\cov
	\,&=\,
		0
	\,,\quad
	\dpb{Q}{\bQ}_\cov
	\,=\,
		-2\mem \bpsi_m\mem \Ric^m{}_k\mem \psi^k
\,,\\[0.1\baselineskip]
	\dpb{Q}{H}_\cov
	\,&=\, 
		- \tfrac{8}{3}\mem 
			\bigbig{ i\hem \bpsi D^m R_\wrap{mk} \psi }
			\mem \psi^k
		+ \tfrac{4}{3}\mem
			p_m \Ric^m{}_k\mem \psi^k
\,,\\[0.125\baselineskip]
	\label{grav.H~H}
	\dpb{H}{H}_\cov
	\,&=\,
\begin{aligned}[t]
	&{
		- \tfrac{4}{3}\mem 
			\bigbig{ i\hem \bpsi D^m R_{mk} \psi }
			\mem p^k
		+ \tfrac{2}{3}\mem
			p_m p_n \Ric^{mn}
	}
	\\[-0.1\baselineskip]
	&
		-
			\bpsi_m \psi^n
		\mem
		\BB{
			D^2 R^m{}_n{}^r{}_s
			+ 2\mem R^m{}_i{}^r{}_j R^i{}_n{}^j{}_s
			- 2\mem R^m{}_i{}^j{}_s R^i{}_n{}^r{}_j
			- R^m{}_n{}^i{}_j R^j{}_i{}^r{}_s
		}\mem
			\bpsi_r \psi^s
	\,,
\end{aligned}
\end{align}
\end{subequations}
where $\Ric_{ns} := R^m{}_{nms}$.
The details are shown below:
\begin{subequations}
\begin{align}
	\dpb{Q}{Q}_\cov
	\,&=\,
\begin{aligned}[t]
	&+ \frac{1}{2}\mem
		\Ric_{rk}\mem \psi^r \psi^k
	\contx{
		\P }{ \X
	}{
		\Ps }{ \bPs
	}\,\\
	&- \frac{1}{2}\mem
		\Ric_{rk}\mem \psi^r \psi^k
	\contxd{
		\X }{ \P
	}{
		\bPs }{ \Ps
	}\,,
\end{aligned}\\
	\dpb{Q}{\bQ}_\cov
	\,&=\,
\begin{aligned}[t]
	&+
		\bigbig{ i\hem \bpsi R_{mn} \psi }\mem (-i\hem\eta^{mn})
	\contx{
		\P }{ \P
	}{
		\Ps }{ \bPs
	}\\
	&-
		\frac{1}{2}\mem \bpsi_m \Ric^m{}_k\mem \psi^k
	\contx{
		\X }{ \P
	}{
		\Ps }{ \bPs
	}\\
	&-
		\frac{1}{2}\mem \bpsi_m \Ric^m{}_k\mem \psi^k
	\contxd{
		\P }{ \X
	}{
		\Ps }{ \bPs
	}\,,
\end{aligned}
	\,,\\
	\dpb{Q}{H}_\cov
	\,&=\,
	\begin{aligned}[t]
		&{}
		\BB{
			- \bigbig{ i\hem \bpsi D^m\hnem R_{mk} \psi }\mem \psi^k
			+ \Ric_{nk}\mem \psi^n p^k
		}
	\contx{
		\P }{ \X
	}{
		\Ps }{ \bPs
	}\\
		&{}
		- \frac{2}{3}\mem 
			\bigbig{ i\hem \bpsi D^m R_\wrap{mk} \psi }
			\mem \psi^k
		+ \frac{1}{3}\mem 
			p_m
			\Ric^m{}_k\mem
			\psi^k
	\contd{
		\X }{ \P
	}{
		\X }{ \P
	}\,,
	\end{aligned}\\
	\label{grav.H0~H0}	
	\dpb{H_0}{H_0}_\cov
	\,&=\,
	\begin{aligned}[t]
		&{}
		\bb{
			- \frac{2}{3}\mem 
				\bigbig{ i\hem \bpsi D^m R_{mk} \psi }
				\mem p^k
			+ \frac{1}{3}\mem
				p_m p_n
				\Ric^{mn}
		}
		\conty{
			\X }{ \P
		}{
			\X }{ \P
		}\\
		&{}
			+
				\bigbig{ i\hem \bpsi R^{ij} \psi }
				\bigbig{ i\hem \bpsi R_{ij} \psi }
		\contd{
			\P }{ \P
		}{
			\P }{ \P
		}\,,
	\end{aligned}
	\\
	\label{grav.H0~H1}
	\dpb{H_0}{H_1}_\cov
	\,&=\,
	\begin{aligned}[t]
		&{}
			- \frac{1}{2}\mem
				\bigbig{\hem\bpsi
					D^2 R^r{}_s
				\psi}
				\mem \bpsi_r \psi^s
		\contd{
			\P }{ \X
		}{
			\P }{ \X
		}\,,
	\end{aligned}
	\\
	\label{grav.H1~H1}
	\dpb{H_1}{H_1}_\cov
	\,&=\,
	\begin{aligned}[t]
		&{}
			\bpsi_m \psi^n\mem R^m{}_i{}^j{}_n
			R^i{}_s{}^r{}_j\mem \psi^s \bpsi_r
		\contx{
			\Ps }{ \bPs
		}{
			\bPs }{ \Ps
		}\\
		&{}
			+ \bpsi_m \bpsi_n R^m{}_i{}^n{}_j
			R^i{}_r{}^j{}_s \psi^r \psi^s
		\contyd{
			\Ps }{ \bPs
		}{
			\Ps }{ \bPs
		}\,.
	\end{aligned}
\end{align}
\end{subequations}

Lastly, 
brute-force calculation
shows 
$R^A{}_{BCD}[\nabla] \neq 0$
and
$\Ric[\nabla]_{AB} \neq 0$.
Remarkably, however,
the nonvanishing components of the Ricci tensor are only
$\Ric[\nabla]_{\X_m \X_n} = \tfrac{2}{3}\mem \Ric_{mn}$,
implying
$\Ric[\nabla]_{AB} = 0$
on spacetime Ricci flatness.

\section{Color-Kinematics Duality and Its Covariant Cousin}
\label{CCK}

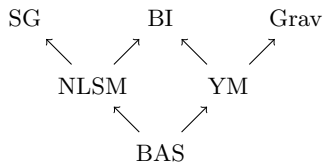
\begin{figure}
\adjustbox{valign=c}{\begin{tikzpicture}
    \node[empty] (O) at (0,0) {};
    \node[empty] (a) at (-0.9,0.9) {};
    \node[empty] (b) at ( 0.9,0.9) {};
    \node[w] (00) at ($(O)$) {\clap{BAS}};
    \node[w] (1a) at ($(O)+1*(a)$) {\clap{NLSM}};
    \node[w] (1b) at ($(O)+1*(b)$) {\clap{YM}};
    \node[w] (2aa) at ($(O)+2*(a)$) {\clap{SG}};
    \node[w] (2ab) at ($(O)+1*(a)+1*(b)$) {\clap{BI}};
    \node[w] (2bb) at ($(O)+2*(b)$) {\clap{Grav}};
    \node[w] (ph00) at ($(O)$) {};
    \node[w] (ph1a) at ($(O)+1*(a)$) {};
    \node[w] (ph1b) at ($(O)+1*(b)$) {};
    \node[w] (ph2aa) at ($(O)+2*(a)$) {};
    \node[w] (ph2ab) at ($(O)+1*(a)+1*(b)$) {};
    \node[w] (ph2bb) at ($(O)+2*(b)$) {};
    \draw[<-] (ph2aa)--(ph1a) node[] {};
    \draw[<-] (ph1a)--(ph00) node[] {};
    \draw[<-] (ph2ab)--(ph1b) node[] {};
    \draw[->] (ph1a)--(ph2ab) node[] {};
    \draw[->] (ph00)--(ph1b) node[] {};
    \draw[->] (ph1b)--(ph2bb) node[] {};
\end{tikzpicture}}
	\caption{
		The web of field theories exhibiting color-kinematics duality.
		The arrows pointing to the left replace $\su(N)$ with $\sdiff(\R^d)$.
		The arrows pointing to the right replace $\su(N)$ with $\g_\YM$, a mystery infinite-dimensional Lie algebra.
	}
	\label{web}
\end{figure}

\para{Review of Color-Kinematics Duality and Its Working Definition}%
Color-kinematics duality
is a remarkable property of
scattering amplitudes 
that establishes a precise correspondence between 
perturbative
gauge theory and gravity.
It has origins in open-closed duality in string theory \cite{KLT}
and has been formulated within quantum field theory 
by Bern, Carrasco, and Johansson (BCJ) \cite{BCJ1, BCJ2},
establishing that 
gauge theory amplitudes ``square'' to
gravitational amplitudes.
This squaring relation has been also observed 
at the level of 
exact
classical solutions,
such as
black holes
\cite{monteiro2014black,Luna:2015paa,nja}
and
gravitational instantons
\cite{Berman:2018hwd,note-sdtn}.
See \rcite{BCJReview} for a comprehensive review.

A working definition of color-kinematics duality
is given in terms of
Bi-Adjoint Scalar (BAS)
theory.
BAS theory is a field theory of a scalar field $\Phi^{a\ta}$
that carries
two Lie algebra indices $a = 1,\cdots,\dim\g$ and $\ta = 1,\cdots,\dim\tg$.
The two Lie algebras $\g$ and $\tg$
are independent,
for which the Jacobi identities must be strictly satisfied by definition.
The equations of motion of the BAS field are
\begin{align}
	\label{BAS}
	\Box\mem \Phi^{a\ta}
	\,=\,
		-
		f^a{}_{bc}\mem 
		\tf^\ta{}_{\tb\tc}\,
			\Phi^{b\tb}\mem \Phi^{c\tc}
	\,.
\end{align}
A theory exhibits
color-kinematics duality
at tree level
if its equations of motion can be formulated as
the BAS equations of motion in \eqref{BAS}
for a choice of the Lie algebras $\g$ and $\tilde{\g}$
\cite{cck}.

Well-established instances are 
Non-Linear Sigma Model (NLSM) and a special instance of Galileon theory---%
Special Galileon (SG) in short---%
in general $d$ dimensions,
which take
$\g {\,=\,} \su(N)$, $\tg {\,=\,} \sdiff(\R^d)$
and
$\g {\,=\,} \tg {\,=\,} \sdiff(\R^d)$,
respectively \cite{cck}.
Here, $\sdiff(\mathbb{R}^d)$ denotes the Lie algebra of volume-preserving diffeomorphisms in $d$ dimensions,
which is infinite-dimensional.
When examined in the Fourier (plane-wave) basis,
it takes momenta as indices
and thus is referred to as a kinematic algebra.

For gauge theories and gravity,
our current understanding on color-kinematics duality
has been far less complete,
and an explicit identification of the kinematic algebra
has been limited to the self-dual sector in four dimensions
\cite{DonalSDYM1,DonalSDYM2,DonalSDYM3,HenrikSDYM,park1990self,husain1994self,plebanski1994self,plebanski1975some}.
Hence, the question of the kinematic algebra in general $d$ dimensions remains unresolved.
As shown in \fref{web},
the relations found from
scattering amplitudes
assert that
YM theory is a BAS theory with $\g = \su(N)$ and $\tg = \g_\YM$,
BI theory is a BAS theory with $\g = \sdiff(\R^d)$ and $\tg = \g_\YM$,
and 
the NS-NS sector of type II supergravity
(including general relativity as a subsector)
is a BAS theory with $\g = \g_\YM$ and $\tg = \g_\YM$.
Here, $\g_\YM$ is a mystery Lie algebra that will be infinite-dimensional.
$\g_\YM$ is the kinematic algebra of YM theory.

\skip
\para{Covariant Color-Kinematics Duality for YM Theory}%
A partial progress has been made by Cheung and Mangan \cite{cck},
where the prototype theory
is taken as 
BAS theory coupled to a gauge connection $A^a{}_\a$:
\begin{align}
	\label{GBAS}
	\eta^{rs} D_r D_s \Phi^{a\ta}
	\,=\,
		- f^a{}_{bc}\mem \tf^\ta{}_{\tb\tc}\,
			\Phi^{b\tb}\mem \Phi^{c\tc}
	\quad\text{where}\quad
	D_r \Phi^{a\ta}
	\,=\,
		\partial_r \Phi^{a\ta}
		+ f^a{}_{cb}\mem A^c{}_r\mem \Phi^{b\ta}
	\,.
\end{align}
In this construction,
the two Lie algebras play asymmetric roles:
the Lie algebra $\g$ is gauged (color index),
whereas
the Lie algebra $\tilde{\g}$ is global (flavor index).
\eqref{GBAS} is referred to as the gauged BAS equations
and is identified as the template for a covariant cousin of color-kinematics duality
by Cheung and Mangan \cite{cck}.

YM theory 
can be formulated
in terms of the gauged BAS equations 
for $\tg = \so(1,d-1)$
\cite{cck}.
To show this, \rcite{cck} implements the following gymnastics of covariant derivatives
on the field strength:
\begin{align}
\begin{split}
	\label{CCK.YM}
	\eta^{rs} D_r D_s F^a{}_{mn}
	\,=\,
		D^r D_r F^a{}_{mn}
	\,&=\,
		- D^r D_m F^a{}_{nr}
		+ D^r D_n F^a{}_{mr}
	\quad\text{by \eqref{YM.1m}}
	\,,\\
	\,&=\,
		- f^a{}_{bc}\mem F^b{}^r{}_m F^c{}_{nr}
		+ f^a{}_{bc}\mem F^b{}^r{}_n F^c{}_{mr}
	\quad\text{by \eqref{YM.1e}}
	\,,\\
	\,&=\,
		-2\mem f^a{}_{bc}\mem F^b{}_{mr} F^{cr}{}_n
	\,=\,
		- f^a{}_{bc}\mem \BB{
			F^b{}_{mr} F^{cr}{}_n
			- F^c{}_{mr} F^{br}{}_n
		}
	\,.
\end{split}
\end{align}
In the second line, we have commuted the covariant derivatives
to convert
$[D^r,D_m]$ and $[D^r,D_n]$
into the field strengths.
This derives the covariant second-order field equations of YM theory,
previously presented in \eqref{YM.2}.
By identifying the antisymmetrized pair of indices $[mn]$ with the Lie algebra index $\ta$ for $\tg = \so(1,d-1)$,
\eqref{CCK.YM} is rewritten as
the gauged BAS equations of motion in \eqref{GBAS}
for $\g = \su(N)$ and $\tg = \so(1,d-1)$.

In the same way,
BI theory
can be formulated
in terms of the gauged BAS equations 
for $\tg = \sdiff(\R^d)$.
This applies the color-to-diffeomorphism replacement
(NLSM replacement rule in \rcite{cck})
to \eqref{CCK.YM}.

In the phase space formulation pursued in the main article,
these gauged BAS equations arise from the following equation 
that follows from combining \eqref{!magnetic} and \eqref{!electric}:
\begin{align}
	\label{!cck}
	\pb{\pb{\pb{p_m}{p_n}}{p_r}}{p^r}
	+ 2\mem \pb{
		\pb{p_m}{p_r}
	}{
		\pb{p^r}{p_n}
	}
	\,=\,
		0
	\,.
\end{align}

\skip
\para{Covariant Color-Kinematics Duality for Gravity}%
Finally, it remains to elaborate on general relativity.
First of all, the true, algebraic statements about the Riemann tensor
in general relativity are
\begin{subequations}
\label{true-GR}
\begin{align}
\label{true-GR.a}
	R_{\m\n\r\s} 
	+ R_{\m\r\s\n}
	+ R_{\m\s\n\r}
	\,&=\,
		0
	\,,\\
\label{true-GR.b}
	\Ric^\m{}_\n
	\,:=\,
		R^{\m\r}{}_{\n\r}
	\,&=\,
		0
	\,,\\
\label{true-GR.c}
	R_{\m\n\r\s} 
	\,&=\,
		R_{\r\s\m\n}
	\,.
\end{align}
\end{subequations}
The first equation is the algebraic Bianchi identity.
The second equation is the Ricci flatness,
implied by the vacuum Einstein's equations.
The last equation states the index symmetry of the Riemann tensor.
Second of all,
the magnetic- and electric-type first-order equations 
can be identified as \cite{cck}
\begin{align}
	\label{GR.1m}
	\nabla_\wrap{[\k} R^{\m\n}{}_\wrap{\r\s]}
	\,&=\,
		0
	\quad\text{(magnetic)}
	\,,\\
	\label{GR.1e}
	\nabla^\r R^{\m\n}{}_{\r\s}
	\,&=\,
		0
	\quad\text{(electric)}
	\,.
\end{align}
\eqref{GR.1m} is the differential Bianchi identity.
\eqref{GR.1e} is implied by
the algebraic Bianchi identity, the Ricci flatness, the index symmetry,
and the differential Bianchi identity:
\begin{align}
	\text{
		\eqref{true-GR} and \eqref{GR.1m}
	}
	\qiq
	\nabla^\r R^{\m\n}{}_{\r\s}
	\,&=\,
		- \nabla^\m R^{\n\r}{}_{\r\s}
		- \nabla^\n R^{\r\m}{}_{\r\s}
	\,=\,
		\nabla^\m \Ric^\n{}_\s
		- \nabla^\n \Ric^\m{}_\s
	\,=\,
		0
	\,.
\end{align}
With this understanding,
we consider the following gymnastics:
\begin{align}
	\label{CCK.GR}
	g^{\k\l}\mem \nabla_\k \nabla_{\nem\l}
		R^\m{}_\n{}^\r{}_\s
	\,&=\,
		\nabla^\l
		\BB{
			 \nabla_{\nem\s} R^\m{}_\n{}^\r{}_\l
			 - (\r \leftrightarrow \s)
		}
	\quad\text{by \eqref{GR.1m}}
	\,,\\
	\,&=\,
		\BB{
		 	  R^\m{}_\k{}^\l{}_\s R^\k{}_\n{}^\r{}_\l
  		 	- R^\m{}_\k{}^\r{}_\l R^\k{}_\n{}^\l{}_\s 
		 	+ R^\r{}_\k{}^\l{}_\s R^\m{}_\n{}^\k{}_\l
			- R^\m{}_\n{}^\r{}_\k R^\k{}_\l{}^\l{}_\s
		}
		 - (\r \leftrightarrow \s)	
	\quad\text{by \eqref{GR.1e}}
	\nonumber
	\,,\\
	\,&=\,
		2\mem \BB{
 	  		R^\m{}_\k{}^\l{}_\s R^\k{}_\n{}^\r{}_\l
		 	- R^\m{}_\k{}^\r{}_\l R^\k{}_\n{}^\l{}_\s 
		}
		 	+ R^\l{}_\k{}^\r{}_\s R^\m{}_\n{}^\k{}_\l
	\quad\text{by \eqrefs{true-GR.a}{true-GR.b}}
	\nonumber
	\,,\\
	\,&=\,
		- 2\mem \BB{
 	  		R^\m{}_\k{}^\r{}_\l R^\k{}_\n{}^\l{}_\s 
		 	- R^\m{}_\k{}^\l{}_\s R^\k{}_\n{}^\r{}_\l
		}
		 	+ R^\m{}_\n{}^\k{}_\l R^\l{}_\k{}^\r{}_\s
	\,.
	\nonumber
\end{align}
Here, the indices are raised and lowered via the metric.
\eqref{CCK.GR}
is known as
the Penrose wave equation
\cite{Penrose:1960eq,Ryan:1974nt}.
By transitioning to the orthonormal frame
via vielbein $E^\m{}_m$,
\eqref{CCK.GR} boils down to
\begin{align}
	\eta^{kl}
		D_k D_l R^m{}_n{}^r{}_s
	\,=\,
		- 2\mem \BB{
 	  		R^m{}_k{}^r{}_l R^k{}_n{}^l{}_s 
		 	- R^m{}_k{}^l{}_s R^k{}_n{}^r{}_l
		}
		 	+ R^m{}_n{}^k{}_l R^l{}_k{}^r{}_s
	\,.
\end{align}
By identifying the antisymmetrized pair of indices $[mn]$ with the Lorentz Lie algebra index,
\eqref{CCK.GR} could be viewed as
an instance of a fully gauged BAS equation
with $\g = \tg = \so(1,d-1)$:
\begin{align}
	\label{GGBAS}
	\eta^{kl} D_k D_l
	\Phi^{\ta_1\ta_2}
	\,=\,
		-\tf^{\ta_1}{}_{\tb_1\tc_1}\mem \tf^{\ta_2}{}_{\tb_2\tc_2}\,
			\Phi^{\tb_1\tb_2}\mem \Phi^{\tc_1\tc_2}
		- 2\mem \Phi^{\ta_1\td}\mem \Phi_\td{}^{\ta_2}
	\quad\text{where}\quad
	D_r \phi^{\ta}
	\,=\,
		E^\r{}_r\hem
		\BB{
			\partial_\r \phi^{\ta}
				+ f^\ta{}_{\tc\tb}\mem A^\tc{}_\r\mem \phi^{\tb}
		}
	\,.
\end{align}
Here, all indices are subject to gauge transformations,
being coupled to the spin connection as a Lorentz-valued gauge connection $A^\ta{}_\r$.
In addition,
\eqref{GGBAS} differs from \eqref{GBAS}
also by the fact that
the covariant derivatives
are dressed with the vielbeins.
Hence the covariant derivative in \eqref{GGBAS}
does not map to the covariant derivative in \eqref{GBAS}
by the mere replacement of $\so(1,d-1)$ with $\su(N)$.
Note also that the term
$2\mem \Phi^{\ta_1\td}\mem \Phi_\td{}^{\ta_2}$
ruins index factorization.

\section{Born-Infeld Theory as Teleparallel Gravity}

In this last appendix, we reproduce Section 5 of 
Mason and Newman \cite{mason-newman}
in our phase space language.
We clarify that the resulting teleparallel theory
shall be identified as
Born-Infeld (BI) theory,
based on the modern reformulation established by Cheung and Mangan \cite{cck}.
We plan to provide
a more detailed investigation on this equivalence
in a future work
\cite{bipaper?}.

\skip
\para{YM Theory from Colored Scalar Particle}%
Recall the Feynman derivation for YM theory,
presented in the main article.
We took $T^*\R^d {\,\times\,} \g^*$ as a Poisson manifold
and considered 
a generic modification of its Poisson structure
that preserves
the spacetime Poisson-commutativity 
as well as the color Lie algebra:
\begin{align}
\begin{split}
	\label{re-YM.feynbr}
	\pb{x^m}{p_n}
	\,=\,
		\delta^m{}_n
	\,,\quad
	\pb{q_a}{q_b}
	\,=\,
		q_c\mem f^c{}_{ab}
	\,,\quad
	\pb{q_a}{p_m}
	\,=\,
		q_b\mem f^b{}_{ca}\mem A^c{}_m(x)
	\,,\quad
	\pb{p_m}{p_n}
	\,=\,
		q_a\mem F^a{}_{mn}(x)
	\,.
\end{split}
\end{align}
Specifically,
define the Jacobiator as
$\Jac{f,g,h} := \pb{\pb{f}{g}}{h} + \pb{\pb{g}{h}}{f} + \pb{\pb{h}{f}}{g}$.
The Feynman derivation
imposes vanishing of Jacobiators.
This 
derives the Jacobi identity of the color Lie algebra $\g = \su(N)$,
the definition of the nonabelian field strength,
and the magnetic first-order equations:
\begin{align}
\label{re-YM.qqq}
	\Jac{q_a,q_b,q_c} = 0
	&\qiq
		f^a{}_{eb}\mem f^e{}_{cd}
		+ f^a{}_{ec}\mem f^e{}_{db}
		+ f^a{}_{ed}\mem f^e{}_{bc}
	\,=\, 0
	\,,\\
\label{re-YM.fsdef}
	\Jac{q_a,p_m,p_n} = 0
	&\qiq
		F^a{}_{mn}
		\,=\,
			\partial_m A^a{}_n - \partial_n A^a{}_m
			+ f^a{}_{bc}\mem A^b{}_m\mem A^c{}_n
	\,,\\
\label{re-YM.1m}
	\Jac{p_m,p_n,p_r} = 0
	&\qiq
		D_\wrap{[r} F^a{}_\wrap{mn]}
		\,=\,
			0
	\,,\\
\label{re-YM.1e}
	\pb{\pb{p_m}{p_n}}{p^n} = 0
	&\qiq
		D^n F^a{}_{mn}
		\,=\,
			0
	\,.
\end{align}

\skip
\para{BI Theory from Scalar Particle}%
To derive gravity in the same fashion,
a natural attempt
is to take 
an even simpler phase space:
$T^*\R^d$.
The free theory features the canonical Poisson brackets
between position and momentum.
In the interacting theory,
the most general modification of the Poisson structure
preserving the Poisson-commutativity of spacetime
is given as
$\pb{x^\m}{p_n} = \delta^\m{}_n \mplus A^\m{}_n$
and $\pb{p_m}{p_n} \neq 0$.
With hindsight, we have employed
another suite of indices
$\m,\n,\cdots$
that also runs through $d$ integers.

With such modified brackets,
$\Jac{x^\m,x^\n,p_r}$
trivially vanishes 
if the deformation $A^\m{}_m$
is a sole function of $x$.
Provided we adopt this assumption,
\smash{$\Jac{x^\m,p_r,p_s} = 0$}
implies that $\pb{p_r}{p_s}$
is at most linear in the momentum.
Since the part independent of the momentum
simply turns out to implement electromagnetic interactions,
we work with the following refined prescription:
\begin{align}
\begin{split}
	\label{BI.feynbr}
	\pb{x^\m}{p_n}
	\,=\, 
		\delta^\m{}_n + A^\m{}_n(x)
	\,=:\,
		E^\m{}_n(x)
	\,,\quad
	\pb{p_m}{p_n}
	\,=\,
		-p_k\mem \Omega^k{}_{mn}(x)
	\,.
\end{split}
\end{align}

Given \eqref{BI.feynbr},
the vanishing of $\Jac{x^\m,p_m,p_n}$ and \smash{$\Jac{p_m,p_n,p_r}$} implies
\begin{align}
\label{BI.fsdef}
	\Jac{x^\m,p_m,p_n} = 0
	&\qiq
	[E_m , E_n]^\m
	\,=\, \Omega^k{}_{mn}\mem E^\m{}_k
	\,,\\
\label{BI.1m}
	\Jac{p_m,p_n,p_r} = 0
	&\qiq
	[ E_\wrap{[r} , \Omega^k{}_\wrap{mn]}\mem E_k ]^\m
	\,=\, 0
	\,.
\end{align}
Here, the bracket denotes the Lie bracket between vector fields:
$[V,W]^\m = V^\n \partial_\n W^\m - W^\n \partial_\n V^\m$.
Geometrically, 
we can interpret $E^\m{}_m\mem \partial_\m$ 
as a set of vector fields in $\R^d$,
or a vielbein in short.
Then
\eqref{BI.fsdef}
implies that $\Omega^k{}_{mn}$
are the anholonomy coefficients
of the vielbein.
Plugging this in,
\eqref{BI.1m}
is trivialized due to the Jacobi identity
of the diffeomorphism Lie algebra
$\diff(\R^d) \cong \Gamma(T\R^d)$.
Observe the parallel between
\eqrefs{BI.feynbr}{re-YM.feynbr},
\eqrefs{BI.fsdef}{re-YM.fsdef},
and
\eqrefs{BI.1m}{re-YM.1m}.

The above construction
achieves the coupling of the particle to a field theory of a vielbein $E^\m{}_m$,
which takes the form of the usual minimal matter coupling.
For instance,
consider 
the Hamiltonian equations of motion (geodesic equation)
or the Lagrangians
facilitated by the vanishing Jacobians.
As before, however,
the dynamics of this vielbein field theory has not been fully specified.
To this end, we take 
the Mason-Newman postulate in \eqref{!electric}
and find
\begin{align}
	\label{BI.1e}
	[ E^n , \Omega^k{}_\wrap{mn}\hem E_k ]^\m
	\,=\,
		0
	\,,
\end{align}
where the internal indices $m,n,\cdots$ are raised and lowered 
via a flat metric $\eta_{mn}$.
By using \eqref{BI.fsdef},
this equation is fully expanded out as
\begin{align}
	\label{BI.1e*}
	E^{\m n}\mem \partial_\m \Omega^k{}_{mn}
	+ \Omega^{kn}{}_l\mem \Omega^l{}_{mn}
	\,=\,
		0
	\,.
\end{align}
Notably,
\eqrefs{BI.fsdef}{BI.1e*} exactly reproduce
{Eqs.}\,(11) and (5.1a) of Mason and Newman \cite{mason-newman}.

Unfortunately, 
it is explicitly addressed in
Section 5 of \rcite{mason-newman}
that this vielbein field theory fails to 
describe
general relativity.
Its interpretation
is rather left unclear,
although
it is remarked that
\eqrefs{BI.fsdef}{BI.1e*} arise in 
an alternative theory of gravity
built by Einstein
during his late-stage research on 
a unified field theory based on absolute parallelism
\cite{cartan1979letters}:
the anholonomy coefficients $\Omega^m{}_{rs}$
are the vielbein-frame components of the teleparallel torsion
due to the Weitzenb\"ock connection.

From a modern perspective, however,
we clarify
that
this vielbein theory
shall be identified as
BI theory in a disguise.
To see this,
convert one index of the anholonomy coefficients
to a spacetime index:
\begin{align}
	\label{FEW}
	F^\m{}_{rs}
	\,:=\,	
		E^\m{}_m\mem \Omega^m{}_{rs}
	\,.
\end{align}
Then,
given the expansion
$E^\m{}_m = \delta^\m{}_m + A^\m{}_m$
around a flat background,
\eqref{BI.fsdef} translates to
\begin{align}
	\label{BI.fsdef(cj)}
	F^\m{}_{rs}
	\,=\,
		\partial_r A^\m{}_s
		- \partial_s A^\m{}_r
		+ [ A_r , A_s ]^\m
	\,,
\end{align}
where $\partial_r := \delta^\r{}_r\mem \partial_\r$.
Accordingly, the magnetic and electric equations in Eqs.\,(\ref{BI.1m}) and (\ref{BI.1e})
boil down to
\begin{align}
	\label{BI.1(cj)}
	D_\wrap{[r} F^\m{}_\wrap{mn]}
	\,=\, 0
	\,,\quad
	D^n F^\m{}_\wrap{mn}
	\,=\,
		0
	\,.
\end{align}
\eqrefs{BI.fsdef(cj)}{BI.1(cj)}
are exactly
the new formulation of BI theory
provided by Cheung and Mangan \cite{cck}.

While
the original formulation of BI theory \cite{born1935quantization}
utilizes an abelian connection as the field basis,
the new formulation due to \rcite{cck}
employs a 
diffeomorphism-valued connection $A^\m{}_m$,
which is a strict implication of color-kinematics duality.
That is, $\g = \diff(\R^d)$ is taken as a gauge algebra.
With this understanding, 
\eqref{BI.1(cj)} has denoted
\begin{align}
	\label{D=LE}
	D_r F^\m{}_{mn}
	\,=\,
		\partial_r F^\m{}_{mn}
		+ [A_r , F_{mn}]^\m
	\,=\,
		[ E_r , F_{mn} ]^\m
	\,.
\end{align}
The field redefinition relating the original and new formulations of BI theory
could be similar to the semiclassical limit of Seiberg-Witten map
\cite{Seiberg:1999vs}.

Interestingly,
the formulation of BI theory in terms of the dynamical vielbein $E^\m{}_m$
is background-independent and also diffemorphism invariant.
In this context, 
the volume-preserving condition on the vielbein
which \rcite{mason-newman}
imposes
could be taken as a diffeomorphism gauge-fixing condition.

To summarize,
we have derived BI theory
as a YM theory for 
$\g = \diff(\R^d)$
by imposing
the Jacobi identity
and the Mason-Newman postulate
on a scalar particle
coupled to a vielbein field,
expanded around a flat background.

Amusingly,
we might learn that
Einstein was quietly
stepping along the path toward double copy
in his late quest for a unification of gravity and gauge theory.

\skip
\para{Teleparallel Torsion Formulation}%
In the above analysis, the indices $m,n,r,s,\cdots$ 
have been global Lorentz indices
just as in YM theory.
An optional pathway, however, is to exploit a Lorentz-valued flat connection
to gauge the indices $m,n,r,s,\cdots$.
This introduces more gauge redundancies
but emphasizes the teleparallel interpretation.

This idea can be approached by examining the symplectic form
giving rise to the brackets in \eqref{BI.feynbr}:
\begin{align}
	\label{BI.symp}
	\vartheta \,=\, p_m\mem e^m
	\qiq
	\omega \,=\, d\vartheta
	\,=\,
		dp_m \swedge e^m
		+ p_m\mem de^m
	\,.
\end{align}
The anholonomy coefficients arise from the well-known identity
\begin{align}
	[E_r , E_s] \,=\, \Omega^m{}_{rs}\mem E_m
	\quad\iff\quad
	de^m + \tfrac{1}{2}\, \Omega^m{}_{rs}\mem e^r \swedge e^s
	\,=\, 0
	\quad\iff\quad
	\Omega^m{}_{rs}
	\,=\,
		-2\mem \gamma^m{}_\wrap{[rs]}
	\,,
\end{align}
where $\gamma^m{}_{nr}$ encodes the Levi-Civita connection in the orthonormal frame:
the spin connection usual in general relativity.
Now, let $\tilde{D}$ be a flat Lorentz-valued connection,
which will be referred to as the teleparallel connection.
Then the symplectic form in \eqref{BI.symp} can be alternatively computed as
\begin{align}
	\label{BI.symp-T}
	\vartheta \,=\, p_m\mem e^m
	\qiq
	\omega \,=\, d\vartheta
	\,=\,
		\tilde{D}p_m \swedge e^m
		+ p_m T^m
	\,,
\end{align}
where the torsion two-form is given by
\begin{align}
	\label{tele-T}
	T^m \,=\, \tilde{D}e^m
	\,=\,
		de^m + \tilde{\gamma}^m{}_{nr}\mem e^r \swedge e^n
	\quad\iff\quad
	T^m{}_{rs}
	\,=\,
		- \Omega^m{}_{rs} - 2\mem \tilde{\gamma}^m{}_{[rs]}
	\,=\,
		2\mem \delta{\gamma}^m{}_{[rs]}
	\,.
\end{align}
Here, $\delta{\gamma}^m{}_{rs} := \gamma^m{}_{rs} - \tilde{\gamma}^m{}_{rs}$
describes the difference between the Levi-Civita and flat connections,
i.e., inertial force as a tensor \cite{wald2010general}.
In fact, by deriving the Hamiltonian equations of motion
as the Hamiltonian flow of $p^2/2$,
one can realize that
the term $p_m\mem T^m$ in \eqref{BI.symp-T}
precisely realizes the notion of gravitational force
in the gravitoelectromagnetism \cite{wald2010general} sense:
compare \eqref{BI.symp-T} with the symplectic form of the colored scalar particle,
$\omega = dp_m \wedge dx^m + i\mem D\btheta_i \wedge D\theta + q_a F^a$,
where $q_a F^a$ implements the nonabelian Lorentz force.
Hence $q_a \leftrightarrow p_m$ and $F^a \leftrightarrow T^m$.
Note also that 
utilizing the Levi-Civita connection for computing the symplectic form
does not generate such a ``force'' term to be single-copied.

By using the teleparallel connection $\tilde{D}$ and the teleparallel torsion $T$ in \eqref{tele-T},
the first-order equations of BI theory
in \eqrefs{BI.1m}{BI.1e}
become
\begin{align}
	\label{BI.1tele}
	\mathfrak{D}_\wrap{[r} T^k{}_\wrap{mn]} = 0
	\,,\quad
	\mathfrak{D}^n T^k{}_{mn} = 0
	\,,
\end{align}
where we have denoted
$
	\mathfrak{D}_r T^k{}_{mn}
	=
		\tilde{D}_r T^k{}_{mn}
		+ T^k{}_{lr}\mem T^l{}_{mn}$.
This shows that the magnetic-type equations for BI theory are nothing but
the Bianchi identity for the teleparallel torsion.
Combining the two equations in \eqref{BI.1tele}, we also obtain
the covariant color-kinematics duality equations for BI theory:
\begin{align}
	\label{CCK.BI(vielbein)}
	\eta^{rs}\mem 
		\mathfrak{D}_r \mathfrak{D}_s\mem T^k{}_{mn}
	\,=\,
		-2\, \BB{
			T^i{}_{rm}\mem \mathfrak{D}_i T^{kr}{}_n
			- T^{ir}{}_{n}\mem \mathfrak{D}_i T^{k}{}_{rm}
			+ T^k{}_{ij}\mem T^i{}_{mr}\mem T^{jr}{}_n
		}
	\,,
\end{align}
where
$
	\mathfrak{D}_r \mathfrak{D}_s T^k{}_{mn}
	=
		\tilde{D}_r ( \mathfrak{D}_s T^k{}_{mn} )
		+ T^k{}_{lr}\mem ( \mathfrak{D}_s T^l{}_{mn} )
$.


	\twocolumngrid
	\bibliography{references.bib}
	
\end{document}